\documentclass[10pt,a4paper]{article}

\usepackage[english]{babel}
\usepackage[utf8]{inputenc} 
\usepackage[inner=3cm,outer=3cm,bottom=3cm,top=3cm]{geometry}
\usepackage{graphicx}
\usepackage{array}
\usepackage{morefloats}
\usepackage{amsfonts,amsmath,color}										
\usepackage{tikz}
\usepackage[T1]{fontenc}    
\usepackage{appendix}
\usepackage{stackrel}
\usepackage{eurosym}
\usepackage{subfigure}
\usepackage{epstopdf} 
\usepackage{amscd,amssymb,amsthm,latexsym, amsbsy,amsmath}
\usepackage{mathtools}
\usepackage{multicol}
\usepackage{multirow}
\usepackage{booktabs}
\usepackage{float}
\usepackage{colortbl}
\usepackage{xcolor}
\usepackage{enumerate} 
\usepackage[round]{natbib} 
\usepackage{rotating}  
\usepackage{textcomp} 
\usepackage{longtable} 
\usepackage{authblk} 
\usepackage{wasysym}
\usepackage{arydshln} 
\usepackage{adjustbox} 
\usepackage{rotating} 
\usepackage{blkarray} 
\usepackage{caption} 
\usepackage[colorlinks=true, linkcolor=blue, citecolor=blue, urlcolor=blue]{hyperref}
\usepackage{tikz}
\usetikzlibrary{fit,shapes.geometric}
\usepackage{soul} 
\usepackage{dashbox}
\usepackage{calc}
\usepackage{ifthen}

\definecolor{bluegray}{rgb}{0.4, 0.6, 0.8} 
\definecolor{darkcerulean}{rgb}{0.03, 0.27, 0.49}
\definecolor{darkpastelgreen}{rgb}{0.01, 0.75, 0.24} 
\definecolor{byzantine}{rgb}{0.74, 0.2, 0.64} 
\definecolor{brightlavender}{rgb}{0.75, 0.58, 0.89} 
\definecolor{coralred}{rgb}{1.0, 0.25, 0.25} 
\definecolor{lightcoral}{rgb}{0.94, 0.5, 0.5}
\definecolor{darkpastelgreen}{rgb}{0.01, 0.75, 0.24} 
\definecolor{limegreen}{rgb}{0.2, 0.8, 0.2}
\definecolor{richelectricblue}{rgb}{0.03, 0.57, 0.82} 
\definecolor{bluegray}{rgb}{0.4, 0.6, 0.8} 
\definecolor{mydeepskyblue}{rgb}{0.0, 0.7, 0.75} 
\definecolor{darkturquoise}{rgb}{0.0, 0.81, 0.82} 
\definecolor{mediumturquoise}{rgb}{0.28, 0.82, 0.8}
\definecolor{turquoise}{rgb}{0.19, 0.84, 0.78}
\definecolor{junglegreen}{rgb}{0.16, 0.67, 0.53} 
\definecolor{junebud}{rgb}{0.74, 0.85, 0.34} 
\definecolor{lapislazuli}{rgb}{0.15, 0.38, 0.61} 
\definecolor{veronica}{rgb}{0.63, 0.36, 0.94} 
\definecolor{unitednationsblue}{rgb}{0.36, 0.57, 0.9} 
\definecolor{lightgray}{gray}{0.9} 
\definecolor{my_seagreen}{rgb}{0, 0.75, 0.77}
\definecolor{blue-green}{rgb}{0.0, 0.87, 0.87}

\newcolumntype{T}[1]{>{\centering\let\newline\\\arraybackslash\hspace{0pt}}m{#1}}
\LTcapwidth=\textwidth

\providecommand{\keywords}[1]{\textbf{\textit{Key words---}} #1}

\DeclareCaptionType{equ}[Matrix][]
\newcolumntype{C}[1]{>{\centering\let\newline\\\arraybackslash\hspace{0pt}}m{#1}}

\newcounter{nodemarkers}
\newcommand\squaretext[1]{%
	\tikz[overlay,remember picture] 
	\node (marker-\arabic{nodemarkers}-a) at (0,1.5ex) {};%
	#1%
	\tikz[overlay,remember picture]
	\node (marker-\arabic{nodemarkers}-b) at (0,0){};%
	\tikz[overlay,remember picture,inner sep=2pt]
	\node[draw,dashed,rectangle,fit=(marker-\arabic{nodemarkers}-a.center) (marker-\arabic{nodemarkers}-b.center), inner sep=3pt] {};%
	\stepcounter{nodemarkers}%
}

\newcommand{\hlc}[2][yellow]{{%
		\colorlet{foo}{#1}%
		\sethlcolor{foo}\hl{#2}}%
}

\title{Guidelines for LASSO and derivatives use under different dependence and scale structures}
\vspace{0.1cm}
\author[1,3]{Laura Freijeiro-González}
\author[2,3]{Manuel Febrero-Bande}
\author[2,3]{Wenceslao González-Manteiga}

\affil[1]{Department of Statistics and Operational Research and Didactics of Mathematics; Universidad de Oviedo (UNIOVI), Oviedo (Asturias), Spain. Email: freijeirolaura@uniovi.es.
	
Orcid: https://orcid.org/0000-0003-1640-9102.}

\affil[2]{CITMAga; Universidade de Santiago de Compostela (USC), Santiago de Compostela (Galicia), Spain.}

\affil[3]{Department of Statistics, Mathematical Analysis and Optimization; Universidad de Santiago de Compostela (USC), Santiago de Compostela (Galicia), Spain.}
\date{}                     
\begin{document}
	

\maketitle

\begin{abstract}

In a multivariate linear regression model with $p>1$ covariates, implementation of penalization techniques often implies a preliminary univariate standardization step. Although this prevents scale effects on the covariates selection procedure, possible dependence structures can be disrupted, leading to wrong results. This is particularly challenging in high-dimensional settings where $p \geq n$. In this paper, we analyze the standardization effect on the LASSO for different dependence-scales contexts by means of an extensive simulation study. Two distinct objectives are pursued: adequate covariate selection and proper predictive capability. Additionally, its behavior is compared with the one of some well-known or innovative competitors. This comparison is also extended to three real datasets facing different dependence-scales patterns. Eventually, we conclude with discussion and guidelines on the most suitable methodology for each case in terms of covariates selection or prediction.
\end{abstract}	



\section{Introduction}\label{intro}

In a multivariate regression setup with $p>1$ covariates, it is expected that a bunch of the $X=(X_1,\dots,X_p)\in\mathbb{R}^p$ considered covariates for the explanation of $Y\in\mathbb{R}$ are not relevant. Thus, these should be avoided in the fitting procedure. As a result, it is advisable to implement a preliminary covariates selection step to reduce the number of noisy features entering the model. This is especially remarkable in high-dimensional regimes where the number of covariates $p$ is greater than the sample size $n$ and classical estimation procedures fail.

In terms of covariates selection, one can modestly classify the existing methodologies in three non-exclusive groups: i) tools trusting in an underlying structure between $X$ and $Y$, ii) thresholding techniques establishing a relevance order between variables and iii) coefficients for measuring some general notion of a certain type of dependence. Examples of the group i) are backward/forward algorithms (see, for example, \cite{Pope1972}, \cite{Hocking1983} or \cite{Draper1998}) or boosting techniques in regression models (see \cite{Friedman2000} or \cite{Friedman2001} to say a few). Related to the second group ii), we can see screening procedures, which sort out covariates and apply some threshold to detect the relevant variables (we refer to \cite{Jianqing2008} or \cite{buhlmann2011statistics}, among others). Finally, block iii) corresponds to new coefficients derived from the kernel based distance covariance ideas of \cite{Gretton2005} and \cite{Szekely2007}. The choice of methodology should align with one's objectives, as certain methods are more suitable for specific goals. For example, the class i) needs some model structure assumption and proper estimation. This translates in the best possible selection of covariates for that given model structure. However, if this hypothesis is incorrect, the variable $Y$ would be badly explained by the covariates in $X$. This would also lead to poor prediction results. In contrast, screening techniques related to group ii) do not require of model estimation to detect relevant terms. These only need a proper tool to rank covariates concerning some dependence strategy (e.g. using some correlation coefficient) and to obtain an appropriate cutoff (e.g. based on some sample quantile). Nonetheless, these procedures inherited the disadvantages of the employed correlation coefficients, jointly with the difficulty of obtaining an optimal threshold in practice. Last, the dependence coefficients of block iii) allow for detecting important covariates in terms of some general dependence concept (e.g. conditional dependence in mean). Again, it is not necessary to assume any structure or fit a regression model. One can check if some covariates, $X$, and the response, $Y$, have some type of dependence relation, resorting to hypothesis testing or some threshold, without knowing what type of relation is established (e.g. linear, additive, etc.). A common issue with blocks ii) and iii) techniques is the loss of interpretability in the performed selection. While these procedures can identify relevant terms, the relationship between covariates and the response remains unclear, not allowing prediction. In consequence, a second specification step would be needed for this purpose. 

In a $p>n$ high-dimensional regime, some extra problems regarding covariates selection techniques arise. Now, it is not possible to resort to type i) techniques as classical approaches to fit regression models fail in this setup. One solution is to incorporate some penalization term in the estimation process. An example is the LASSO regression of \cite{Tibshirani1996}, which adds an $L_1$-type penalization to the linear regression model formulation. Concerning class ii) of covariates selection algorithms, we can not apply similar ideas as in the $n\geq p$ case as classical correlation coefficients do not work properly in the $p>n$ framework. Furthermore, prediction is not available in this context for procedures of type ii) or iii). Roughly speaking, we can not apply some second specification step given that classical goodness of fit tests fail in the $p>n$ framework.

In view of the previous limitations, it seems reasonable to assume a certain structure between both variables, $Y$ and $X$, and work under this premise, especially in the $p>n$ high-dimensional framework. This will allow one not to only select covariates, but to establish a relation between selected explanatory terms and response to perform prediction a posteriori. In particular, we are going to focus on the easiest structure: the linear regression model formulation. Then, one can resort to modified type i) techniques for the $p>n$ case, as the penalizations approach. We will apply and analyze different penalization techniques using the well-known LASSO approach as the core of this study. Later, we will compare their performance with competitors following blocks i) and iii) ideas, as well as a comparison with procedures applying ii) philosophy.

The LASSO regression is a penalization technique introduced by \cite{Tibshirani1996}. This approach proposes the imposition of a $L_1$ penalization in a linear regression model, performing covariates selection and overcoming the drawback of the $\beta$ estimation in the $p>n$ high-dimensional framework. This estimator results in
\begin{equation}
	\label{bLASSO} 
	\hat{\beta}^{LASSO} \coloneqq \mathop{\arg \min}\limits_{\beta} \left\lbrace \sum_{i=1}^{n}\left(y_i-\sum_{j=1}^{p}x_{ij}\beta_j\right)^2+ \lambda\sum_{j=1}^{p}|{\beta_j}| \right\rbrace,
\end{equation}
where $\{(x_i,y_i),i=1,\dots,n\}$ is an independent and identically distributed (idd) sample from the joint distribution function of $(X,Y)\in\mathbb{R}^p\times\mathbb{R}$, $\beta\in\mathbb{R}^p$ and $\lambda>0$ is the penalty parameter. We assume henceforth and without loss of generality that Y and X are centered variables.

The resulting $\hat{\beta}^{LASSO}$ of (\ref{bLASSO}) is a sparse vector, getting some null terms. For large values of $\lambda>0$, the coefficients of $\beta$ are more penalized, which results in a higher number of elements that are shrinkaged to zero. As a result, one can apply both simultaneously: $\beta$ estimation and covariates selection in the $p>n$ regime. This last can be performed just selecting those covariates which associated $\hat{\beta}^{LASSO}_j\neq 0$, for $j=1,\dots,p$.

The LASSO problem has been widely studied over the last years owing to its good statistical properties. See, for example, the review of \cite{tibshirani2011regression}. It has been showed that this procedure is consistent in terms of prediction (see \cite{van2009conditions} for an extensive analysis), and this guarantees consistency of the parameter estimates at least in a $L_2$ sense (\cite{van2009conditions}, \cite{meinshausen2009lasso}, \cite{candes2007dantzig}); besides, this is a consistent variable selector under some theoretical conditions (\cite{Meinshausen2006}, \cite{wainwright2009sharp}, \cite{Zhao2006}).

In spite of all these good qualities, the LASSO regression has some important limitations in practice (see for example \cite{Zou2005} or \cite{Su2017}). These are related to its biased nature (see Chapter 3 of \cite{hastie2009elements}, Chapter 4 of \cite{giraud2014introduction} or Chapter 2 of \cite{hastie2015statistical}), the great number of false discoveries (see \cite{wasserman2009high} and \cite{Su2017}), and its difficulty in estimating a proper value of the regularization parameter $\lambda$ in practice (see \cite{Yang2005}, \cite{Leng2006}, \cite{Meinshausen2006}, \cite{buhlmann2011statistics}, \cite{Lahiri2021}, \cite{Dalalyan2017} and \cite{homrighausen2018study}). Furthermore, an additional limitation is the requirement of strict conditions in the model design for proper covariates recovering (see \cite{Meinshausen2006}, \cite{Zhao2006}, \cite{Zou2006}, \cite{Yuan2007}, \cite{Bunea2008}, \cite{Meinshausen2010} and \cite{buhlmann2011statistics}). These conditions can not be always verified in practice, specifically when there exist strong dependence patterns between covariates, resulting in possible collinearity effects. Related to this last, \cite{Zou2005} proved that, when high-dependence structures exit, i.e. some covariates highly correlated, the LASSO algorithm tends to pick some of them randomly and avoid the remaining ones. This selection could result in a loss of information if these related terms are relevant or in a confusion phenomenon when there are strong relations between important and noisy covariates. We refer to \cite{FreijeiroGonzalez2022} for an extensive review of these topics.

Apart from the concern about dependence structures, one must consider the effect of covariates on different scales. In practice, dependence scenarios often involve covariates with a wide range of values. Some real data examples verifying these conditions are displayed in Section \ref{data}. As a result, it is interesting to determine if the scale effect of the covariates has a role in the LASSO selection and prediction capability. For example, to know if there is an increment of the confusion phenomenon, or a decrease of the prediction capability, when there are noisy covariates with higher/lower scales than relevant ones under dependence frameworks. To the best of our knowledge, no existing literature addresses this topic in the LASSO framework. As a result, in this work, we analyze the dependence and scale effects jointly. For this aim, we test the performance of the LASSO procedure under different dependence-scales structures by means of a broad simulation study. Besides, we compare its performance with suitable competitors. We consider two cases: results selecting covariates using the raw data and employing the classical LASSO approach standardizing these first in a univariate manner. Later, these procedures are tested in three real datasets. 

The paper is organized as follows. Section \ref{scales_LASSO} provides a discussion about the scales effects on the LASSO procedure. Next, the different scenarios considered in the simulation study are introduced in Section \ref{simulation_scenarios}. In Section \ref{LASSO_results}, the LASSO performance is tested in practice. These results are compared with the ones of some important competitors in the literature of similar nature and some thresholding techniques in Section \ref{competitors}. It follows the Section \ref{data} with applications of all considered methodologies to three real datasets. Eventually, some discussion and guidelines for proper covariates selection and prediction accuracy arise in Section \ref{discussion}.


\section{Scales effects on LASSO}\label{scales_LASSO}

In terms of the LASSO problem displayed in (\ref{bLASSO}), one can see as the penalty parameter $\lambda>0$ is the same for all the $\beta_j$ coefficients associated to the $X_j$ covariates, for $j=1,\dots,p$. Thus, all covariates are constrained by the same quantity $\lambda>0$ regardless their scales. In particular, if we can assume that the sample matrix $\bold{X}\in\mathbb{R}^{n\times p}$ is orthonormal\footnote{$\bold{X}$ has orthogonal columns, i.e. linearly independent covariates, with unit norm.}, we can see in \cite{Tibshirani1996}, \cite{hastie2009elements} or \cite{buhlmann2011statistics} that $\hat{\beta}^{LASSO}= \text{sign}(\hat{\beta}_j)\left( |\hat{\beta}_j|-\lambda \right)_+$,  where $\hat{\beta}_j$ is the ordinary least squares (OLS) estimator and $(\cdot)_+=\max\{0,\cdot\}$. This implies that all coefficients are shrinkaged by the same quantity $\lambda>0$, if $|\hat{\beta}_j|>\lambda$, or forced to be null otherwise.

Related to the fact that the same penalty value $\lambda>0$ is considered for all covariates in the LASSO procedure, although their relevance or scale, other approaches have been proposed along the literature to solve this issue. Some proposals adapted to the $p>n$ framework are the Adaptive LASSO (AdapL) approach of \cite{Zou2006} or the SLOPE criterion of \cite{Bogdan2015}. In the first case, the AdapL technique proposes to apply data-dependent weights in the penalization part of (\ref{bLASSO}). Then, one considers $\lambda_j=\lambda w_j$, where $\lambda>0$ and the $w_j>0$ values collect some information about relevance or scale of the $X_j$ covariates, for $j=1,\dots,p$. Some options for these weights can be seen in \cite{Zou2006}. In contrast, the SLOPE procedure considers values $\lambda_1\geq\dots\geq \lambda_p\geq0$ for penalizing the $X_{(1)},\dots,X_{(p)}$ associated covariates. Here, the $X_{(j)}$ terms are the covariates sorted out in decreasing order of relevance regarding to the model, for $j=1,\dots,p$. We refer the reader to \cite{Bogdan2015} for more in-depth details. As a result, these procedures provide adapted weights for each covariate. Nevertheless, the proper selection of the terms $w_j$ in the AdapL procedure or the establishment of the order of relevance for covariates in the SLOPE criterion are tricky problems in practice, especially in the $p>n$ context.

On the other hand, keeping just the orthogonal assumption of $\bold{X}\in\mathbb{R}^{n\times p}$, it can be proved that 
\begin{equation}\label{bLASSO_orthog}
	\hat{\beta}^{LASSO}_j=\frac{1}{\sigma_j^2}\bold{\widetilde{X}}_j^\top Y \left( 1-\frac{\lambda}{2 \sigma_j|\bold{\widetilde{X}}_j^\top Y|} \right)_+  \quad \text{for} \; j=1,\dots,p,
\end{equation}
where $\sigma_j^2>0$ is the $j^{th}$ element of the resulting diagonal matrix $\bold{X}^\top \bold{X}$, concerning the variance of $\bold{X}_j$ when the covariates are centered, and $\bold{\widetilde{X}}_j=\bold{X}_j/\sigma_{j}$ is the univariate standardization version of $\bold{X}_j$, for $j=1,\dots,p$. Hence, in the independence framework, the LASSO approach selects only those covariates verifying $|\bold{\widetilde{X}}_j^\top Y |>\lambda/2\sigma_j$. We refer to Chapter 4 of \cite{giraud2014introduction} for guidelines about calculation of (\ref{bLASSO_orthog}) and more insight for the orthonormal framework.

Paying attention to the condition $|\bold{\widetilde{X}}_j^\top Y |>\lambda/2\sigma_j$ derived from equation (\ref{bLASSO_orthog}) in the orthogonal setting, one can notice the important role of the scale effect in the LASSO selection procedure. Covariates with larger scales, i.e., higher associated values $\sigma_j > 0$, are more likely to exceed the $\lambda/2\sigma_j$ term and therefore be selected by the model. Just considering a raw covariate $\bold{X}_j$ and its univariate standardization version $\bold{\widetilde{X}}_j$, the associated conditions for a nonnull $\hat{\beta}^{LASSO}_j$ coefficient are $|\bold{\widetilde{X}}_j^\top Y |>\lambda/2\sigma_j$ and $|\bold{\widetilde{X}}_j^\top Y |>\lambda/2$, respectively. Thus, the raw version has a higher probability of being incorporated to the model for values $\sigma_{j}>1$, although both provide the same information of the model response. This can lead to a confusion phenomenon, as irrelevant covariates with large scales could be selected while important ones with smaller scale values are avoided. 

Conversely, in a non-orthogonal case, i.e. when there exists dependence relations between covariates, the previous formula for $\hat{\beta}^{LASSO}$ displayed in (\ref{bLASSO_orthog}) does not apply. In particular, it is only possible to know that $\hat{\beta}^{LASSO}$ is not completely null when $\lambda<2\|\bold{X}^\top Y \|_\infty$, where $\|\bold{X}^\top Y\|_\infty=\sup_{j}|\bold{X}_j^\top Y|$ is the infinity vectorial norm, for $j=1,\dots,p$. However, it is no possible to obtain an explicit expression for $\hat{\beta}^{LASSO}$. This translates in not knowing what covariates are selected by the LASSO procedure. See, for example, Chapter 4 of \cite{giraud2014introduction} for more details about this topic.

Concerning the confusion phenomenon, a preliminary standardization step could be convenient. In practice, it is quite common to work with standardized variables in the LASSO procedure when the features are suspected to be in different units. This translates in considering that the $Y$ and $X$ terms are centered and that the columns of $X$ have unit variance. See for example \cite{Tibshirani1996}, \cite{fan2001variable}, \cite{Leng2006}, \cite{zou2007degrees}, \cite{hastie2009elements}, \cite{tibshirani2011regression}, \cite{buhlmann2011statistics}, \cite{Vidaurre2013}, \cite{hastie2015statistical}, \cite{Su2017} or \cite{Ali2019}, to say a few. In high-dimensional settings where $p>n$, only univariate standardization can be employed, as estimating the covariance matrix structure in the classic way is not possible. This applies for orthogonal designs in $\bold{X}$, but does not when there exists some dependence structures. In this last case, the application of univariate standardization disrupts the dependence relations between the covariates and could lead to wrong selections.

As a result, this confusion problem is stressed when there exists any structure of dependence between the covariates. In those cases, it is not possible to know the explanation capability of each covariate in terms of response, as the dependence pattern is usually unknown and difficult to estimate. Besides, the relevant covariates may not be the highest scale ones but may be quite related to some of them. As a result, it is interesting to know how is the performance of the LASSO technique in these types of situations: if this methodology can correctly detect the important terms, using the dependence structure if convenient, or whether this is blurred by the scale effects. Furthermore, we wonder if it is more convenient to work always with the data standardized in a univariate way, if it makes more sense to keep the raw data and apply data-dependent weights or if this depends on the scenario. We try to bridge this gap analyzing the LASSO performance under different dependence-scales structures. For this purpose, an extensive simulation study is provided. Next, the considered simulation scenarios are introduced.


\section{Simulation scenarios}\label{simulation_scenarios}

Three different simulation scenarios are considered with $p=100$ covariates and sample sizes of $n=25$, $50$, $100$, $150$, $300$. All of these verify the consistent condition of $n>\log(p)s\approx4.61s$ (see \cite{Meinshausen2006}), except for $n=25$ and $n=50$ in some cases. Here, $S$ is the set of true relevant covariates and $s=|S| $ its cardinal. Furthermore, $\beta$ is generated trying to guarantee the signal recovery property of $\inf_{j\in S} \lvert \beta_j \lvert > \sqrt{s\log(p)/n}$ (see \cite{buhlmann2011statistics}). The model error $\varepsilon\sim N(0,\sigma_{\varepsilon}^2)$, where its variance, $\sigma_{\varepsilon}^2$, is calculated to verify that the $90\%$ of deviance can be explained at most\footnote{For example, in the Independence framework (IND) case one obtains $\sigma_{\varepsilon}=\sqrt{ \frac{1-0.9}{0.9} \cdot 10 \cdot (1.25)^2 } \simeq 1.317616$ (see Section A of the Appendix for complete calculations).}. For this study, we carry out a total of $M=500$ Monte Carlo replicates. The selection capability is tested by counting the number of covariates correctly selected ($|\hat{S} \cap S|$) and the noisy ones picked ($|\hat{S}\setminus S|$) over the total ($\hat{S}$). Moreover, the prediction accuracy is measured by computing the mean square error (MSE=RSS/n) and the percentage of explained deviance as $\%Dev=(RSS_0-RSS)/RSS_0$. Here $RSS_0=\sum_{i=1}^{n}(y_i-\overline{y})^2$ and $RSS=\sum_{i=1}^{n}\hat{\varepsilon}_i^2$ is the residual sum os squares of the model. Those cases correcting overestimation and those with an associated MSE in the Good Interval (GI) $[0.9\cdot\text{MSE},1.1\cdot\text{MSE}]$, respectively, are highlighted.

\begin{itemize}
	\item{\textbf{Independence (IND)}. Only the first $s=10$ values are not equal zero for $\beta_j$ ($\beta_1=\dots=\beta_{10}=1.25$), while $\beta_j=0$ for all $j=11,\dots,p$. $X$ is simulated as a $N_n(0,\Sigma_p)$, where the covariance matrix $\Sigma$ has different diagonal structures:
		\begin{itemize}
			\item{\textbf{IND}: we assume all covariates in the same scale taking $\Sigma=I_p$.}
			\item{\textbf{RC.IND} (Relevant Covariates are not standardized): Covariance matrix is given by the structure $\text{diag}\left(\Sigma^{\text{RC.IND}}\right)=(0.5,0.5,1,1,3,3,10,10,25,25;\, 1,\stackbin{p-10)}{\dots},1)$}
			\item{\textbf{RNC.IND} (Relevant and Noisy Covariates not standardized): we add different scales for the next $12$ noisy covariates ($j=11,\dots,22$) in the RC.IND structure. This translates in $\text{diag}(\Sigma^{\text{RNC.IND}})=\left( \left(\text{diag}(\Sigma^{\text{RC.IND}})_j\right)_{j=1}^{10};\, 0.5,0.5,1.5,1.5,3,3,10,10,25,25,50,50;\right.$ \\ $\left. 1,\stackbin{p-10-12)}{\dots},1 \right)$.}
	\end{itemize}}
	
	\item{\textbf{Unitary Toeplitz covariance (UTOEP)}. Again, only $s$ ($p>s>0$) covariates are important, simulating $X$ as a $N_n(0,\Sigma)$ and assuming $\beta_j=0.5$ in the places where $\beta\not=0$. In this case, $\sigma_{jk}=\rho^{|j-k|}$ for $j,k=1,\dots,p$ and $\rho=0.5,0.9$. Now, two different dependence structures varying the location of the s relevant covariates are analyzed:
		\begin{itemize}
			\item{\textbf{UTOEP-B} (Block structure): the relevant covariates are the first $s=15$.}
			\item{\textbf{UTOEP-S} (Sparse structure): consider $s=10$ relevant variables placed every 3 sites, which means that only the $\beta_3,\beta_{6},\beta_{9},\dots,\beta_{30}$ terms of $\beta$ are not null.}
	\end{itemize}}
	
	\item{\textbf{Toeplitz covariance with Sparse structure (TOEP-S)}. Similar structure as UTOEP-S, but adding different covariates scales. We take $\Sigma^{\text{TOEP-S}}=D\cdot\Sigma^{\text{UTOEP}}\cdot D^\top$ with $\rho=0.5,0.9$ and $D$ a diagonal matrix given by $\text{diag}(D)=\left( \sigma_1, \sigma_2, \dots, \sigma_p \right)^\top$:
		\begin{itemize}
			\item{\textbf{RC.TOEP-S} (Relevant Covariates not standardized):  relevant covariates have variance equal to $\text{diag}(\Sigma^{\text{RC.IND}})$. This means $\sigma^2_{3}=0.5, \sigma^2_{6}=0.5, \sigma^2_{9}=1 \dots, \sigma^2_{30}=25$.}
			\item{\textbf{RNC.TOEP-S} (Relevant and Noisy Covariates not standardized): we add noisy covariates with different scales. In particular, $\sigma^2_{2}=0.5$, $\sigma^2_{5}=0.5$, $\sigma^2_{8}=1.5$, $\sigma^2_{11}=1.5$, $\sigma^2_{14}=3$, $\sigma^2_{17}=3$, $\sigma^2_{20}=10$, $\sigma^2_{23}=10$, $\sigma^2_{26}=25$, $\sigma^2_{29}=25$, $\sigma^2_{32}=50$ and $\sigma^2_{35}=50$.}
	\end{itemize}}
	
\end{itemize}

These three scenarios have distinct dependence structures with varied scales on the covariates. We start analyzing the performance in the easiest context concerning dependence structure and possible confusion phenomena: the orthogonal framework. Here, there are three different configurations of the covariates' scales: all covariates in the same scale (IND), only some relevant covariates have different magnitudes (RC.IND), and noisy ones in different scales added as well to the previous model (RNC.IND). Next, in the second scenario, all covariates have unit variance and are related through a Toeplitz covariance matrix (UTOEP), mimicking a functional dependence pattern. This setting is an example where the irrepresentable condition holds (see \cite{buhlmann2011statistics}), but the algorithm suffers from highly correlated relations between the actual set of covariates and unimportant ones. The LASSO performance has been previously tested in this framework: see, for example, \cite{Meinshausen2010} or \cite{buhlmann2011statistics}. As the relevant covariates’ location plays an important role, we consider two configurations: relevant covariates are located in the first $s=15$ places (UTOEP-B) and $s=10$ important covariates spread every three locations (UTOEP-S). Eventually, UTOEP-S is mixed with different scales in the third scenario (TOEP-S). This gives place to new adaptations of UTOEP-S: only important covariates with different scales (RC.TOEP-S) and the previous scenario changing the variance of some noisy terms related to the important ones (RNC.TOEP-S). Note that, when assuming different scales for covariates, we consider amounts less and greater than the unit.

In all these settings, we work with the response $y$ and matrix $\mathbf{X}$ centered. Then, we assume a linear regression model without intercept. We compare two different ways of proceed: working with the raw data ($\mathbf{X}^{\text{r}}$) or applying a univariate standardization by columns ($\mathbf{X}^{\text{us}}$). The $\inf_{j\in S} \lvert \beta_j \lvert > \sqrt{s\log(p)/n}$ condition of \cite{buhlmann2011statistics} is guaranteed for IND, but for $n=25$, and only taking $n=300$ in UTOEP and TOEP-S scenarios.


\section{Standardization effect on LASSO}\label{LASSO_results}

In this section, we analyze the standardization effect on the LASSO for covariates selection under different dependence-scales frameworks. For this goal, simulation scenarios introduced in Section \ref{simulation_scenarios} are employed. Here, only a summary of the results is showed. Complete results are collected in Section D of the Appendix.

We implement the standard LASSO algorithm using the library \texttt{glmnet} (\cite{Friedman2010}) of the R software (\cite{R}). For this aim, we make use of the \texttt{cv.glmnet} and \texttt{glmnet} functions. The function \texttt{cv.glmnet} uses the widely employed K-fold cross-validation approach (CV) to select the $\lambda$ parameter which minimizes the MSE, $\lambda^{\min}$. We denote this procedure by LASSO.min (see \cite{Friedman2010} for more details). In order to be capable of comparing different models and following recommendations of the existing literature, we have fixed $K=10$ for all simulations. We also consider two extra approaches: selecting $\lambda$ as the largest
value verifying that this is within 1 standard error of $\lambda^{\min}$ (LASSO.1se) and selecting the penalization parameter as the value that optimize the BIC criterion (LASSO.BIC). The grid of tuning parameter values is taken of length 100 and is calculated based on the sample data and methodology employed, following author's recommendation. This way of proceed is also followed in the implementation of competitors considered in Section \ref{competitors}. More discussion about this topic can be seen in Section 2.1.4 and Section 3 of the Supplementary material of \cite{FreijeiroGonzalez2022}. Besides, we apply the two-step LASSO-OLS version of \cite{Belloni2013} to adjust the model using the relevant covariates selected by the \texttt{glmnet} function for these $\lambda$ values\footnote{\label{foot2}Code for all simulations and real datasets analysis is available in the public GitHub repository \url{https://github.com/LauraFreiG/Covariates_selection.git}. In particular, this is summarized in the folder ``Different scales and dependence effects scenarios'', inside the folder ``Linear Regression''.}.

\begin{figure}[htb]\centering
	\includegraphics[width=\linewidth]{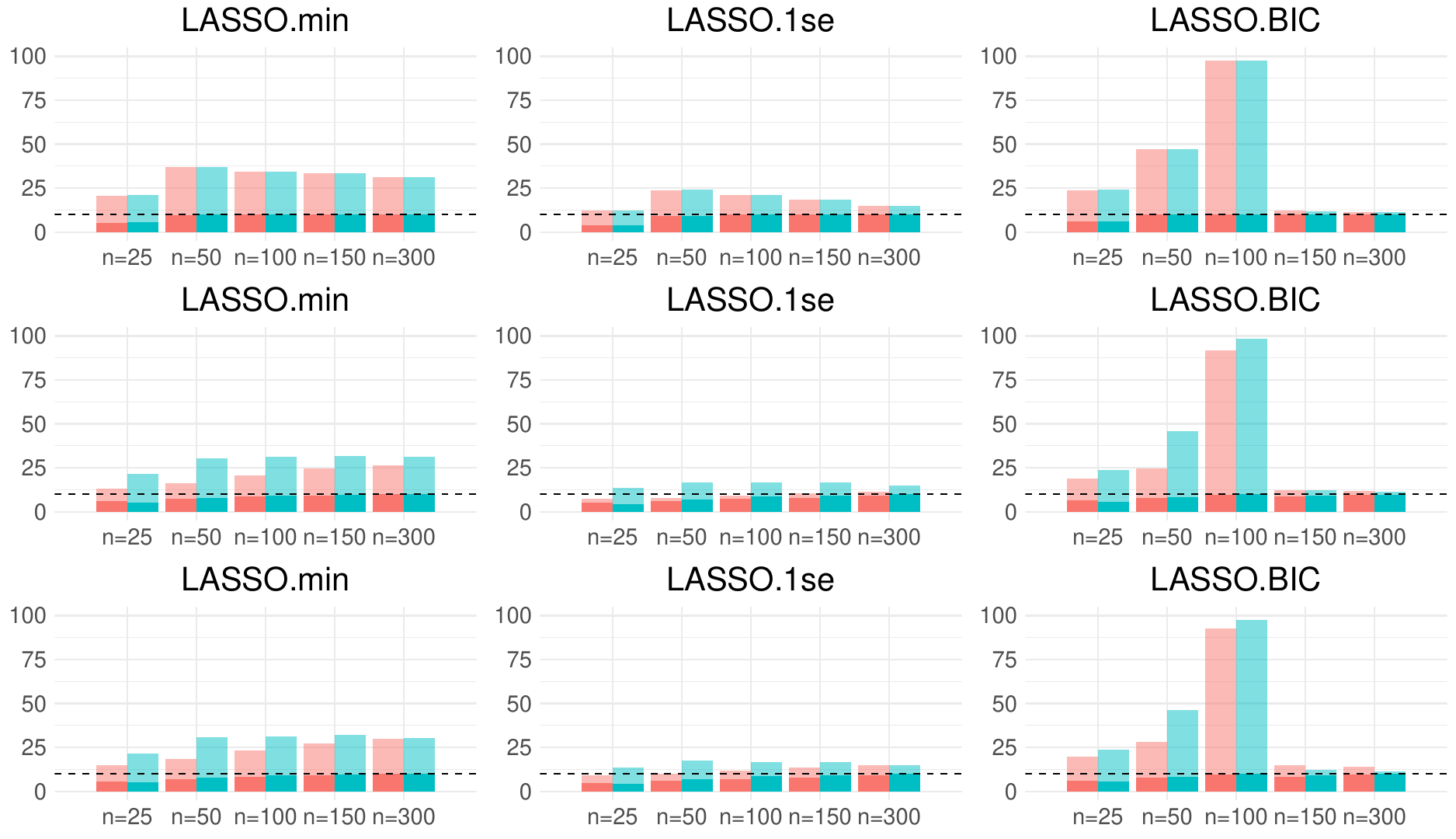}
	\caption{\label{cov_selected_LASSO_1a_1b_1c}Number of important covariates (dark \textcolor{coralred}{left}/\textcolor{my_seagreen}{right} rectangular area) and noisy ones (soft \textcolor{lightcoral}{left}/\textcolor{blue-green}{right} rectangular area) for $p=100$ selected in terms of $\mathcolor{coralred}{\mathbf{X}^{\text{r}}}$/$\mathcolor{my_seagreen}{\mathbf{X}^{\text{us}}}$ in scenarios IND (first row), RC.IND (second row) and RNC.IND (third row). The dashed line marks the $s=10$ value.}
\end{figure}

First, we start analyzing its performance in the easiest framework: orthogonal design. In the IND scenario, one would expect for an efficient algorithm to be able to detect relevant covariates without adding too much noise. Additionally, for RC.IND and RNC.IND, a suitable algorithm is expected not to be influenced by different scales, particularly in RNC.IND, where there are noisy covariates with larger scales than important ones.

 \begin{table}[htb] 
 	\centering
 	\adjustbox{max width=\textwidth}{
 		\begin{tabular}{c cc cc cc cc cc cc}
 			\toprule
 			& \multicolumn{4}{c}{IND} & \multicolumn{4}{c}{RC.IND} & \multicolumn{4}{c}{RNC.IND} \\
 			\cmidrule(rl){2-5} \cmidrule(rl){6-9} \cmidrule(rl){10-13}
 			\rule{0pt}{0.35cm} & \multicolumn{2}{c}{\textbf{RAW}} &  \multicolumn{2}{c}{\textbf{UNIV.}} & \multicolumn{2}{c}{\textbf{RAW}} &  \multicolumn{2}{c}{\textbf{UNIV.}}  & \multicolumn{2}{c}{\textbf{RAW}} &  \multicolumn{2}{c}{\textbf{UNIV.}} \\
 			\cmidrule(r){2-3}  \cmidrule(rl){4-5} \cmidrule(rl){6-7} \cmidrule(rl){8-9} \cmidrule(rl){10-11} \cmidrule(l){12-13}
 			\vspace{-0.07cm}
 			\rule{0pt}{0.35cm} \multirow{2}{*}{ \textbf{METHOD}} & {\footnotesize \textbf{MSE}} &  {\footnotesize \textbf{\% Dev}} & {\footnotesize \textbf{MSE}} &  {\footnotesize \textbf{\% Dev}} & {\footnotesize \textbf{MSE}} &  {\footnotesize \textbf{\% Dev}} & {\footnotesize \textbf{MSE}} &  {\footnotesize \textbf{\% Dev}} & {\footnotesize \textbf{MSE}} &  {\footnotesize \textbf{\% Dev}} & {\footnotesize \textbf{MSE}} &  {\footnotesize \textbf{\% Dev}} \\
 			& {\scriptsize$(1.736)$} & {\scriptsize$(0.9)$} & {\scriptsize$(1.736)$} & {\scriptsize$(0.9)$} & {\scriptsize$(13.715)$} & {\scriptsize$(0.9)$} & {\scriptsize$(13.715)$} & {\scriptsize$(0.9)$} & {\scriptsize$(13.715)$} & {\scriptsize$(0.9)$} & {\scriptsize$(13.715)$} & {\scriptsize$(0.9)$} \\ 
 			\cmidrule(r){1-3} \cmidrule(rl){4-5} \cmidrule(rl){6-7} \cmidrule(rl){8-9} \cmidrule(rl){10-11} \cmidrule(l){12-13}
 			\rule{0pt}{0.35cm} \textbf{LASSO.min} & 1.345 & 0.922 & 1.346 & 0.922 & 10.972 & 0.919 & 10.638 & 0.922 & 10.866 & 0.920 & 10.677 & 0.921 \rule[-0.25cm]{0pt}{0pt} \\
 			\textbf{LASSO.1se} & 1.538 & 0.910 & 1.539 & 0.910 & \cellcolor[gray]{.9}{12.972} & 0.904 & 12.174 & 0.910 & \cellcolor[gray]{.9}{12.813} & 0.891 & 12.180 & 0.891 \rule[-0.25cm]{0pt}{0pt} \\ 
 			\textbf{LASSO.BIC} & \cellcolor[gray]{.9}{1.616} & 0.906 & \cellcolor[gray]{.9}{1.616} & 0.906 & \cellcolor[gray]{.9}{12.671} & 0.907 & \cellcolor[gray]{.9}{12.739} & 0.906 & \cellcolor[gray]{.9}{12.729} & 0.906 & \cellcolor[gray]{.9}{12.722} & 0.906 \rule[-0.25cm]{0pt}{0pt} \\   
 			\bottomrule
 		\end{tabular}
 	}
 	\caption{Results taking $p=100$ and $n=300$ using $\mathbf{X}^{\text{r}}$ and $\mathbf{X}^{\text{us}}$ in IND scenarios. Oracle values are in brackets. \hlc[gray!20]{Highlighted} terms are values in  \small{$[0.9\cdot\text{MSE},1.1\cdot\text{MSE}]$} (GI). }
 	\label{summ_LASSO_scenario_1}
 \end{table}

 A summary of the LASSO results under independence is collected in Figure \ref{cov_selected_LASSO_1a_1b_1c} and Table \ref{summ_LASSO_scenario_1}. Complete results are displayed in Table 16 of Section D.1 of the Appendix. As expected, one can notice a similar behavior between $\mathbf{X}^{\text{r}}$ and $\mathbf{X}^{\text{us}}$ results, although $\mathbf{X}^{\text{us}}$ adds more noise in the selection procedures in RC.IND and RNC.IND scenarios. Concerning covariates selection, the LASSO approaches (LASSO.min, LASSO.1se and LASSO.BIC) always select a number of terms larger than the optimal one $s=10$, adding noise to the model (see Figure \ref{cov_selected_LASSO_1a_1b_1c}). Besides, these completely recover all relevant covariates for $n\geq50$, i.e. when consistent conditions are verified. In the $n>p$ case, the LASSO.BIC is the methodology that obtains the best results. Conversely, this approach performs poorly in the $p\geq n$ regime, as the BIC criterion tends to strongly overfit the results (see for example \cite{Giraud2012}). Additionally, one can notice as the LASSO.min is more conservative in the sense of guaranteeing the maximum recovery of $S$ no matter the noise addition. In contrast, the LASSO.1se selects fewer covariates, losing some of the $S$ terms, but given more guarantees in terms of true positives. Furthermore, if one studies the percentage of times each of the $j=1,\dots,p$ relevant terms are included in the model for $\mathbf{X}^{\text{r}}$ and $\mathbf{X}^{\text{us}}$ taking $n=300$ in RC.IND and RNC.IND scenarios, we can notice as the LASSO selection is influenced by covariates scale effects. We refer to Figures 18 and 19 in Section D.1 of the Appendix for RC.IND and RNC.IND selection, respectively. The three procedures tend to select the covariates with associated higher scales for $\mathbf{X}^{\text{r}}$ ($j=7$, $8$, $9$, $10$ for RC.IND and $j=7$, $8$, $9$, $10$, $17$, $18$, $19$, $20$, $21$ ,$22$ for RNC.IND), no matter if these are relevant or not. This fact is slightly corrected using the $\mathbf{X}^{\text{us}}$ approach. Thus, $\mathbf{X}^{\text{us}}$ seems to protect against false discoveries when there are noisy covariates with greater scales than those associated with important terms in the orthogonal framework. In terms of prediction, the three variants tend to overestimate the results (see Table \ref{summ_LASSO_scenario_1}). Here the value of the mean squared error (MSE) and the percentage of explained deviance (\% Dev) are always smaller and higher than the oracle values, respectively. The LASSO.BIC obtain MSE values in the GI for the three scenarios and the LASSO.1se for RC.IND and RNC.IND using $\mathbf{X}^{\text{r}}$. However, this only applies in the $n=300$ case (see complete results in Table 16 of Section D.1 of the Appendix). Besides, comparing their results, it seems that the LASSO.1se using $\mathbf{X}^{\text{r}}$ tends to obtain the least overestimation.

\begin{figure}[htb]\centering
	\includegraphics[width=\linewidth]{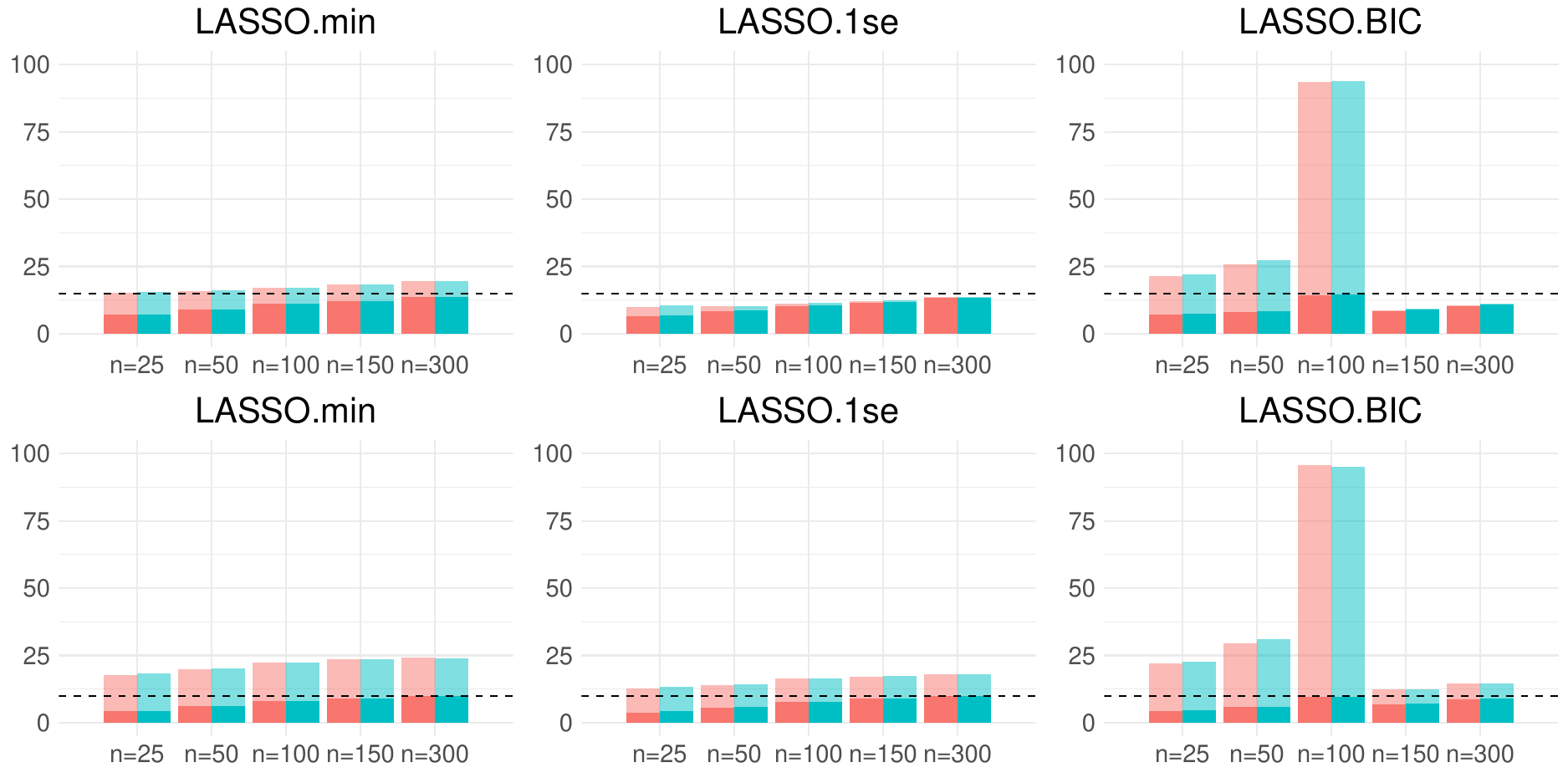}
	\caption{\label{cov_selected_LASSO_2a_2b_rho_9}Number of important covariates (dark \textcolor{coralred}{left}/\textcolor{my_seagreen}{right} rectangular area) and noisy ones (soft \textcolor{lightcoral}{left}/\textcolor{blue-green}{right} rectangular area) for $p=100$ and $\rho=0.9$ selected in terms of $\mathcolor{coralred}{\mathbf{X}^{\text{r}}}$/$\mathcolor{my_seagreen}{\mathbf{X}^{\text{us}}}$ in scenarios UTOEP-B (first row) and UTOEP-S (second row). The dashed line marks the $s=15$ and $s=10$ value for first and second row, respectively.}
\end{figure}

 Next, we test the LASSO performance in a dependence framework: the UTOEP scenarios (UTOEP-B and UTOEP-S). All covariates are assumed to have a Topelitz covariance structure and unit scale. Two different cases of dependence structures are considered: when relevant covariates are the first $s=15$ (UTOEP-B) and when there are only $s=10$ and these are placed every three locations in $j=3,\dots,30$ places (UTOEP-S). In the first scenario (UTOEP-B), the important covariates are highly correlated among them and little with the rest. Particularly, there are only notable confusing correlations in the case of the last variables of $S=\{1,\dots,15\}$ with their noisy neighbors. Here, the LASSO can only recover $S$ in the $n=300$ case due to $\beta$-$\min$ conditions. In contrast, in UTOEP-S, the important covariates are markedly correlated with unimportant ones, so this location magnifies the spurious correlations phenomenon. For this scenario, one also needs a sample size of $n=300$ for proper recovery. LASSO results are summarized in Figure \ref{cov_selected_LASSO_2a_2b_rho_9} and Table \ref{summ_LASSO_scenario_2} for scenarios UTOEP-B and UTOEP-S taking $\rho=0.9$. Complete results taking $\rho=0.5$ and $\rho=0.9$ are collected in Tables 17 and 18 of Section D.2 of the Appendix.

\begin{table}[htb] 
	\centering
	\adjustbox{max width=\textwidth}{
		\begin{tabular}{c cc cc cc cc}
			\toprule
			& \multicolumn{4}{c}{UTOEP-B} & \multicolumn{4}{c}{UTOEP-S} \\
			\cmidrule(rl){2-5} \cmidrule(l){6-9} 
			\rule{0pt}{0.35cm} & \multicolumn{2}{c}{\textbf{RAW}} &  \multicolumn{2}{c}{\textbf{UNIV.}} & \multicolumn{2}{c}{\textbf{RAW}} &  \multicolumn{2}{c}{\textbf{UNIV.}} \\
			\cmidrule(r){2-3}  \cmidrule(rl){4-5} \cmidrule(rl){6-7} \cmidrule(l){8-9} 
			\vspace{-0.07cm}
			\rule{0pt}{0.35cm} \multirow{2}{*}{ \textbf{METHOD}} & {\footnotesize \textbf{MSE}} &  {\footnotesize \textbf{\% Dev}} & {\footnotesize \textbf{MSE}} &  {\footnotesize \textbf{\% Dev}} & {\footnotesize \textbf{MSE}} &  {\footnotesize \textbf{\% Dev}} & {\footnotesize \textbf{MSE}} &  {\footnotesize \textbf{\% Dev}} \\
			& {\scriptsize$(3.807)$} & {\scriptsize$(0.9)$} & {\scriptsize$(3.807)$} & {\scriptsize$(0.9)$} & {\scriptsize$(1.244)$} & {\scriptsize$(0.9)$} & {\scriptsize$(1.244)$} & {\scriptsize$(0.9)$} \\ 
			\cmidrule(r){1-3} \cmidrule(rl){4-5} \cmidrule(rl){6-7} \cmidrule(l){8-9} 
			\rule{0pt}{0.35cm} \textbf{LASSO.min} & \cellcolor[gray]{.9}{3.525} & 0.910 & \cellcolor[gray]{.9}{3.526} & 0.910 & 1.096 & 0.911 & 1.098 & 0.911 \rule[-0.25cm]{0pt}{0pt} \\
			\textbf{LASSO.1se} & \cellcolor[gray]{.9}{3.715} & 0.905 & \cellcolor[gray]{.9}{3.715} & 0.905 & \cellcolor[gray]{.9}{1.154} & 0.906 & \cellcolor[gray]{.9}{1.154} & 0.906 \rule[-0.25cm]{0pt}{0pt} \\
			\textbf{LASSO.BIC} & \cellcolor[gray]{.9}{3.822} & 0.902 & \cellcolor[gray]{.9}{3.800} & 0.903 & \cellcolor[gray]{.9}{1.183} & 0.904 & \cellcolor[gray]{.9}{1.180} & 0.904 \rule[-0.25cm]{0pt}{0pt} \\   
			\bottomrule
		\end{tabular}
	}
	\caption{Results taking $p=100$, $n=300$ and $\rho=0.9$ using $\mathbf{X}^{\text{r}}$ and $\mathbf{X}^{\text{us}}$ in UTOEP scenario. Oracle values are in brackets. \hlc[gray!20]{Highlighted} terms are values in  \small{$[0.9\cdot\text{MSE},1.1\cdot\text{MSE}]$} (GI). }
	\label{summ_LASSO_scenario_2}
\end{table}

Concerning UTOEP-B framework (Table 17 in Section D.2 of the Appendix) we can appreciate differences in terms of the $\rho$ parameter. For $\rho=0.5$ (see Figure 22 in Section D.2 of the Appendix) the three algorithms try to recover the full $S$ set having a similar behavior for $\mathbf{X}^{\text{r}}$ and $\mathbf{X}^{\text{us}}$. They successfully select the $s=15$ relevant covariates under consistent conditions ($n=300$). The LASSO.1se and LASSO.BIC (for $n>p$) obtain the best results, while the LASSO.min adds several noisy covariates in the process. In contrast, taking  $\rho=0.9$ (Figure \ref{cov_selected_LASSO_2a_2b_rho_9}), only the LASSO.min seems to search for the complete recovery of $S$, adding noise in exchange. On the contrary, the LASSO.1se and LASSO.BIC (when $n>p$) seem to make use of the dependence structure, selecting less than $s=15$ covariates and guaranteeing that almost all of them are relevant, i.e. protecting against false discoveries. Paying attention to the number of necessary covariates to explain certain percentage of variability for UTOEP-B (see Table 14 in Section B of the Appendix), one can appreciate that this selection would be also optimal. This can be explained by considering that for $\rho=0.9$, some relevant covariates pick up almost all the information of other nearby ones. Again, a similar behavior is appreciated for $\mathbf{X}^{\text{r}}$ and $\mathbf{X}^{\text{us}}$. Respectively to UTOEP-S, the three procedures try always to recover the full set of relevant covariates, $S$, for $\rho=0.5$ (Figure 22 in Section D.2 of the Appendix) as well as $\rho=0.9$ (Figure \ref{cov_selected_LASSO_2a_2b_rho_9}). Similar to UTOEP-B, $\mathbf{X}^{\text{r}}$ and $\mathbf{X}^{\text{us}}$ behave similarly. Here, one can see as for $\rho=0.9$ (see Figure \ref{cov_selected_LASSO_2a_2b_rho_9}) the procedures interchange relevant terms with some unimportant ones quite correlated due to spurious correlations.

Related to predictions for UTOEP-B and UTOEP-S, we can see similar results to the orthogonal scenario. A summary for $n=300$ and $\rho=0.9$ is displayed in Table \ref{summ_LASSO_scenario_2}. Complete results are collected in Tables 17 and 18 of Section D.2 of the Appendix. Now, we can see that the three approaches keep overestimating the prediction accuracy. Only for $n=300$ all MSE values are in the GI, but for LASSO.min with $\rho=0.5$ and $\rho=0.9$ in UTOEP-S.

\begin{figure}[htb]\centering
	\includegraphics[width=\linewidth]{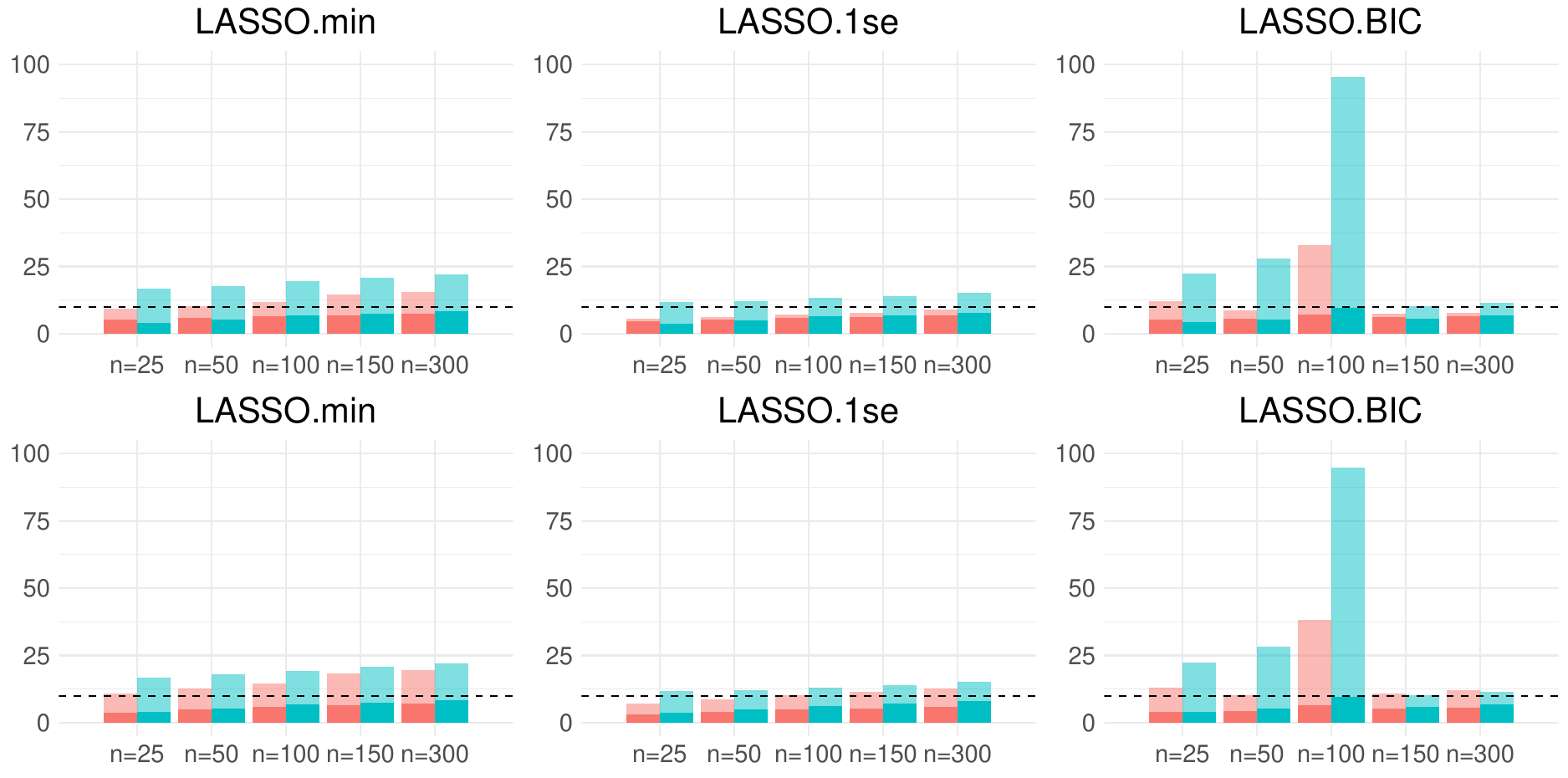}
	\caption{\label{cov_selected_LASSO_3a_3b_rho_0.9}Number of important covariates (dark \textcolor{coralred}{left}/\textcolor{my_seagreen}{right} rectangular area) and noisy ones (soft \textcolor{lightcoral}{left}/\textcolor{blue-green}{right} rectangular area) for $p=100$ and $\rho=0.9$ selected in terms of $\mathcolor{coralred}{\mathbf{X}^{\text{r}}}$/$\mathcolor{my_seagreen}{\mathbf{X}^{\text{us}}}$ in scenarios RC.TOEP-S (first row) and RNC.TOEP-S (second row). The dashed line marks the $s=10$ value.}
\end{figure}

Eventually, we move to a more challenging framework: TOEP-S. We keep the Toeplitz dependence structure of UTOEP-S but allow changes in the scale values of the relevant covariates (RC.TOEP-S), as well as in some irrelevant terms pretty related to the first ones (RNC.TOEP-S). Thus, this is a dependence structure with covariates in different scales, mimicking a real data problem. It is expected for an adequate procedure to be able to use the dependence structure, especially for strong correlations ($\rho=0.9$), and avoid irrelevant ones highly correlated with the important terms. Figure \ref{cov_selected_LASSO_3a_3b_rho_0.9} and Table \ref{summ_LASSO_scenario_3} show a summary of the results for $\rho=0.9$. The remaining results are displayed in Figure 32 and Tables 21 and 22 in Section D.3 of the Appendix.

In terms of TOEP-S scenarios, we can find differences in the selection process related to the scales configuration, the dependence strength and if a preliminary standardization is employed. As expected, there is more noise addition in RNC.TOEP-S framework in comparison to RC.TOEP-S, especially for the $\mathbf{X}^{\text{r}}$ approach. This is explained due to the inclusion of unimportant covariates with high scales. Nonetheless, the $\mathbf{X}^{\text{us}}$ approach tends to add more noise to the model than the raw data one in all these frameworks. Additionally, the three LASSO procedures search for a complete recovery of the $s=10$ relevant covariates in both RC.TOEP-S and RNC.TOEP-S frameworks. However, some noise is included in all cases. When necessary conditions for consistent covariates selection are verified ($n=300$) one can check that these three procedures recover all the relevant terms in the $\rho=0.5$ case (see Figure 31 in Section D.3 of the Appendix). In comparison, for the $\rho=0.9$ configuration (Figure \ref{cov_selected_LASSO_3a_3b_rho_0.9}), some important covariates are interchanged with unrelevant ones quite correlated (see Figure 33 in Section D.3 of the Appendix). Paying attention to the percentage of times each of the relevant covariates is selected (see Figure 32 for RC.TOEP-S and Figure 33 for RNC.TOEP-S in Section D.3 of the Appendix), one can detect differences between $\mathbf{X}^{\text{r}}$ and $\mathbf{X}^{\text{us}}$ selection. Concerning RC.TOEP-S, in all cases, relevant covariates with pretty high scale ($j=21$, $24$, $27$, $30$) are always selected with probability one. Conversely, covariates with lowest scales are selected a less percentage of times, especially when dependence is strong as in the $\rho=0.9$ case. Indeed, $\mathbf{X}^{\text{us}}$ recovers relevant covariates with low scales a percentage of times greater than the one obtained for $\mathbf{X}^{\text{r}}$, but includes more noise as trade-off. A similar behavior is obtained in RNC.TOEP-S. Here, using $\mathbf{X}^{\text{r}}$ translates in a increment of noise in comparison to RC.TOEP-S. This is owing to the inclusion of unimportant covariates with high scales. Analyzing what covariates are selected (see Figure 33 in Section D.3 of the Appendix), one can see as $\mathbf{X}^{\text{r}}$ is easily confused for these, selecting a higher percentage of times covariates $j=32,35$ (especially for $\rho=0.9$). Additionally, this approach has difficulties to detect relevant covariates with low-scales ($j=3,6$ cases). Instead, this procedure selects neighbors quite correlated and with greater scale values. These drawbacks are slightly corrected by applying $\mathbf{X}^{\text{us}}$, although more noisy covariates are added in exchange. As a result, TOEP-S scenarios are examples where the $\mathbf{X}^{\text{r}}$ technique is blurred by the covariates' scales and the $\mathbf{X}^{\text{us}}$ methodology disrupts the dependence structure. Besides, although $\mathbf{X}^{\text{us}}$ seems more ``consistent'' to scale effects, this adds more noise than $\mathbf{X}^{\text{r}}$ for a pretty similar rate of $S$ recovery. An explanation is that, as $\mathbf{X}^{\text{r}}$ tends to select covariates with the highest scales, this selects terms that better explain the data variability. Hence, $\mathbf{X}^{\text{r}}$ requires fewer covariates than $\mathbf{X}^{\text{us}}$, which assumes equal importance of the variables a priori. Besides, $s-2$ of the relevant covariates have scales greater than the unit, being possible candidates for $\mathbf{X}^{\text{us}}$ but not for $\mathbf{X}^{\text{r}}$. In contrast, if the large-scale variables would have only been the unimportant ones, we would expect $\mathbf{X}^{\text{r}}$ not to recover the $S$ set. Opposite, one would expect $\mathbf{X}^{\text{us}}$ to correct this drawback, achieving a better recovery.

\begin{table}[htb] 
	\centering
	\adjustbox{max width=\textwidth}{
		\begin{tabular}{c cc cc cc cc}
			\toprule
			& \multicolumn{4}{c}{RC.TOEP-S} & \multicolumn{4}{c}{RNC.TOEP-S} \\
			\cmidrule(rl){2-5} \cmidrule(l){6-9} 
			\rule{0pt}{0.35cm} & \multicolumn{2}{c}{\textbf{RAW}} &  \multicolumn{2}{c}{\textbf{UNIV.}} & \multicolumn{2}{c}{\textbf{RAW}} &  \multicolumn{2}{c}{\textbf{UNIV.}} \\
			\cmidrule(r){2-3}  \cmidrule(rl){4-5} \cmidrule(rl){6-7} \cmidrule(l){8-9} 
			\vspace{-0.07cm}
			\rule{0pt}{0.35cm} \multirow{2}{*}{ \textbf{METHOD}} & {\footnotesize \textbf{MSE}} &  {\footnotesize \textbf{\% Dev}} & {\footnotesize \textbf{MSE}} &  {\footnotesize \textbf{\% Dev}} & {\footnotesize \textbf{MSE}} &  {\footnotesize \textbf{\% Dev}} & {\footnotesize \textbf{MSE}} &  {\footnotesize \textbf{\% Dev}} \\
			& {\scriptsize$(7.818)$} & {\scriptsize$(0.9)$} & {\scriptsize$(7.818)$} & {\scriptsize$(0.9)$} & {\scriptsize$(7.818)$} & {\scriptsize$(0.9)$} & {\scriptsize$(7.818)$} & {\scriptsize$(0.9)$} \\ 
			\cmidrule(r){1-3} \cmidrule(rl){4-5} \cmidrule(rl){6-7} \cmidrule(l){8-9} 
			\rule{0pt}{0.35cm} \textbf{LASSO.min} & \cellcolor[gray]{.9}{7.100} & 0.908 & 6.920 & 0.910 & 7.022 & 0.909 & 6.926 & 0.911 \rule[-0.25cm]{0pt}{0pt} \\
			\textbf{LASSO.1se} & \cellcolor[gray]{.9}{7.561} & 0.902 & \cellcolor[gray]{.9}{7.302} & 0.906 & \cellcolor[gray]{.9}{7.563} & 0.902 & \cellcolor[gray]{.9}{7.302} & 0.906 \rule[-0.25cm]{0pt}{0pt} \\
			\textbf{LASSO.BIC} & \cellcolor[gray]{.9}{7.643} & 0.901 & \cellcolor[gray]{.9}{7.527} & 0.903 & \cellcolor[gray]{.9}{7.546} & 0.903 & \cellcolor[gray]{.9}{7.526} & 0.903 \rule[-0.25cm]{0pt}{0pt} \\
			\bottomrule
		\end{tabular}
	}
	\caption{Results taking $p=100$, $n=300$ and $\rho=0.9$ using $\mathbf{X}^{\text{r}}$ and $\mathbf{X}^{\text{us}}$ in Scenario 3. Oracle values are in brackets. \hlc[gray!20]{Highlighted} terms are values in  \small{$[0.9\cdot\text{MSE},1.1\cdot\text{MSE}]$} (GI). }
	\label{summ_LASSO_scenario_3}
\end{table}

For prediction in TOEP-S (see Table \ref{summ_LASSO_scenario_3}), similar results as the ones of UTOEP arise. The LASSO techniques keep overestimating the prediction accuracy. Besides, for $n=300$, only LASSO.1se and LASSO.BIC verify that the MSE values are in the GI in all cases. We refer to Tables 21 and 22 in Section D.3 of the Appendix for more results and cases where the LASSO.1se and LASSO.BIC procedures also verify this condition.

\section{Comparison with competitors}\label{competitors}	

Here, we compare LASSO results with adaptations and competitors of this procedure, considering suitable approaches of different nature. In particular, we test the performance of the Adaptive LASSO introduced in \cite{Zou2006} (AdapL.min and AdapL.1se) as a data-dependent weights methodology. Furthermore, we have considered the SCAD procedure of \cite{fan1997comments}, the Dantzig selector approach of \cite{candes2007dantzig} (Dant), the Relaxed LASSO presented in \cite{meinshausen2007relaxed} (RelaxL), the Square root LASSO of \cite{belloni2011square} (SqrtL), the Scaled LASSO of \cite{Sun2012} (ScalL) and the Distance covariance procedure for covariates selection of \cite{febrero2019variable} (DC.VS). A summary of these procedures is displayed in Table \ref{competitors_table}. Additional in-depth details can be found in Section 4 of \cite{FreijeiroGonzalez2022}. Implementation of these algorithms is available in the GitHub repository\footref{foot2}.

\begin{table}[htb] 
	\centering
	\LARGE 
	\setlength{\tabcolsep}{-0.45cm}
	\adjustbox{max width=\textwidth}{
		\begin{tabular}{ccc}
			\toprule
			\rule[-1ex]{0pt}{2.5ex} \textbf{ALGORITHM} & \textbf{METHODOLOGY} & \textbf{IMPLEMENTATION}  \\
			\hline
			\rule[0.4cm]{0pt}{2.5ex} LASSO & \hspace{-0.15cm}\multirow{2}{*}{$\mathop{\min}\limits_{\beta} \left\lbrace \sum_{i=1}^{n}\left(y_i-\sum_{j=1}^{p}x_{ij}\beta_j\right)^2+ \lambda\sum_{j=1}^{p}|{\beta_j}| \right\rbrace$} & \texttt{library(glmnet)}: \Large{\texttt{cv.glmnet}}, \Large{\texttt{glmnet}}  \\
			\rule[-0.1cm]{0pt}{2.5ex} \Large{\cite{Tibshirani1996}} &   & \Large{\cite{Friedman2010}} \rule[-0.55cm]{0pt}{0pt}\\
			\hline
			\rule[0.65cm]{0pt}{2.5ex} AdapL  & \hspace{-0.15cm}$\stackbin[\beta]{}{\min} \left\lbrace \sum_{i=1}^{n}\left(y_i-\sum_{j=1}^{p}x_{ij}\beta_j\right)^2+ \lambda\sum_{j=1}^{p} w_j|{\beta_j}| \right\rbrace$  & \texttt{library(glmnet)}: \Large{\texttt{cv.glmnet}}, \Large{\texttt{glmnet}} \\
			\rule[-0.1cm]{0pt}{2.5ex} \Large{\cite{Zou2006}} & \hspace{-0.15cm}{\Large ($w_j=1/|\hat{\beta}^{RR}_j|^q$ where $\hat{\beta}^{RR}$ is the ridge estimator}  & \Large{\cite{Friedman2010}}\\
			\rule[-0.2cm]{0pt}{2.5ex}  & \hspace{-0.15cm}{\Large of \cite{hoerl1970ridge} and $q\geq1$)}  &  \rule[-0.6cm]{0pt}{0pt}\\
			\hline
			\rule[0.6cm]{0pt}{2.5ex} SCAD & {\Large$\stackbin[\beta]{}{\min} \left\lbrace \sum_{i=1}^{n}\left(y_i-\sum_{j=1}^{p}x_{ij}\beta_j\right)^2+ p_{\lambda}(\beta) \right\rbrace$}  & \texttt{library(ncvreg)}: \Large{\texttt{cv.ncvreg}}, \Large{\texttt{ncvreg}} \\
			\rule[-0.1cm]{0pt}{2.5ex} \Large{\cite{fan1997comments}} & \multirow{3}{*}{\Large $p_{\lambda}(\beta)= \begin{cases}
					\lambda |\beta|, &|\beta|\leq \lambda, \\
					\left(2a\lambda |\beta|-\beta^2-\lambda^2\right)/ \left(2(a-1)\right), &\lambda < |\beta| \leq a\lambda \;\; (a>2) \\
					\left(\lambda^2 (a+1)\right)/2, &\text{otherwise}.
				\end{cases} 
				$}  & \Large{\cite{ncvreg}}\\
			\rule[0cm]{0pt}{2.5ex}  &   &  \\
			\rule[0cm]{0pt}{2.5ex}  &   &  \rule[-0.4cm]{0pt}{0pt}\\	
			\hline
			\rule[0.5cm]{0pt}{2.5ex} Dant & $\stackbin[\beta]{}{\min}  \|\beta\|_1 \; \text{s.t.} \;  \| X^\top r \|_{\infty} \leq \lambda_p \cdot \sigma $  & \texttt{library(flare)}: \Large{\texttt{slim}} \\
			\rule[-0.1cm]{0pt}{2.5ex} \Large{\cite{candes2007dantzig}} & {\Large ( $ \| X^\top r \|_{\infty}:= \stackbin[1\leq j \leq p]{}{\sup} |(X^\top r)_j| $ and $r=y-X \beta$ )}  & \Large{\cite{flare}} \rule[-0.85cm]{0pt}{0pt}\\
			\hline
			\rule[0.5cm]{0pt}{2.5ex} RelaxL & \hspace{1cm}$\stackbin[\beta]{}{\min} \left\lbrace n^{-1}\sum_{i=1}^{n}\left( y_i-x_i^\top \{\beta \cdot \mathbf{1}_{\mathcal{M}_\lambda}\} \right)^2+ \phi \lambda\|\beta\|_1 \right\rbrace$  & \texttt{library(relaxo)}: \Large{\texttt{cvrelaxo}} \\
			\rule[-0.1cm]{0pt}{2.5ex} \Large{\cite{meinshausen2007relaxed}} & where $\phi\in(0,1]$  & \Large{\cite{relaxo}} \rule[-0.5cm]{0pt}{0pt}\\
			\hline
			\rule[0.5cm]{0pt}{2.5ex} SqrtL & \hspace{1.5cm}\multirow{2}{*}{$\stackbin[\beta]{}{\min} \left\lbrace \left[ \sum_{i=1}^{n}\left(y_i-\sum_{j=1}^{p}x_{ij}\beta_j\right)^2\right]^{1/2} + \lambda\sum_{j=1}^{p}|{\beta_j}| \right\rbrace$}  & \texttt{library(flare)}: \Large{\texttt{slim}} \\
			\rule[-0.1cm]{0pt}{2.5ex} \Large{\cite{belloni2011square}} &   & \Large{\cite{flare}} \rule[-0.8cm]{0pt}{0pt}\\
			\hline
			\rule[0.1cm]{0pt}{2.5ex} \multirow{1.5}{*}{ScalL} & \hspace{1cm}\multirow{2}{*}{$\hat{\sigma}\leftarrow \|y-X\hat{\beta}^{old}\|_2/ n^{1/2}, \quad \lambda\leftarrow \hat{\sigma} \lambda_0$} & \multirow{1.5}{*}{\texttt{library(scalreg)}: \Large{\texttt{scalreg}}} \\
			\multirow{1.7}{*}{\Large{\cite{Sun2012}}} & & \multirow{1.7}{*}{ \Large{\cite{scalreg}} } \\
			 & \hspace{1cm}\multirow{2}{*}{$\hat{\beta}^{new} = \stackbin[\beta]{}{\min} \left\lbrace
				\begin{aligned}
					&x^\top_j(y-X\hat{\beta})/n=\lambda \textrm{sign}(\hat{\beta}_j), &\hat{\beta}_j \not=0,\\
					&x^\top_j(y-X\hat{\beta})/n \in \lambda[-1,1], &\hat{\beta}_j =0.\\
				\end{aligned} 
				\right.$} &  \\
			\rule[0cm]{0pt}{2.5ex} &   & \\
			\rule[0.2cm]{0pt}{2.5ex}  & \hspace{1cm}$\hat{\beta}\leftarrow \hat{\beta}^{new}, \; L_{\lambda}(\hat{\beta}^{new}) \leq  L_{\lambda}(\hat{\beta}^{old})$ &  \\
			\rule[0.05cm]{0pt}{2.5ex}  &  \hspace{1.5cm}{\Large $\left(L_{\lambda}(\beta)=(\|y-X\beta\|_2^2)/(2n)+\lambda \sum_{j=1}^{p} | \beta_j | \right)$} &  \rule[-0.75cm]{0pt}{0pt}\\
			\hline
			\rule[0.4cm]{0pt}{2.5ex} DC.VS & $\hat{\varepsilon}\leftarrow y$, \; $M\leftarrow \emptyset$, \; $J\leftarrow \{1,\dots,p\}$  & \texttt{library(fda.usc)}: \Large{\texttt{fregre.glm.vs}} \\
			\rule[-0.1cm]{0pt}{2.5ex} \Large{\cite{febrero2019variable}} & $x_j / DC(\hat{\varepsilon},x_j)\geq DC(\hat{\varepsilon},x_{j'})$ $\forall j, j' \in J$  & \Large{\cite{fda.usc}} \\
			\rule[-0.1cm]{0pt}{2.5ex}  & and $H_0\colon \hat{\varepsilon} \perp x_j$ is rejected:  &  \\
			\rule[-0.1cm]{0pt}{2.5ex}  & \hspace{0.5cm}$M\leftarrow M \cup \{x_j\}$, \; $J\leftarrow J - \{x_j\}, \; \hat{\varepsilon}\leftarrow y- \beta \mathbf{X}_M $  &  \\
			\rule[-0.1cm]{0pt}{2.5ex}  &  \hspace{0.5cm}\Large{(iterate steps 2-3 till step 2 is not verified or $J=\emptyset$)} &  \rule[-0.5cm]{0pt}{0pt}\\
			\bottomrule
		\end{tabular}		
	}
	\caption{Considered algorithms, related methodology and employed library and functions of \texttt{R} (\cite{R}) for practical implementation\footref{foot2}. }
	\label{competitors_table}
\end{table}

Following similar guidelines as the ones presented for the LASSO algorithm (see Section \ref{LASSO_results}), we apply a two-step procedure. First, a covariates selection step is carried out using the different competitors. Next, each corresponding linear model is fitted by minimizing the OLS criterion just with the previously chosen covariates. As a result, the selection capability is analyzed in the first step and the prediction accuracy in the second one. For this purpose, we employ the simulation scenarios introduced above in Section \ref{simulation_scenarios}. Specifically, a distinction between results for the $p>n$ framework (taking $n=50$) and those for the $n>p$ context (considering $n=300$) is made.

\subsection{Context of $\bf p>n$}

\begin{figure}[htb]\centering
	\includegraphics[width=\linewidth]{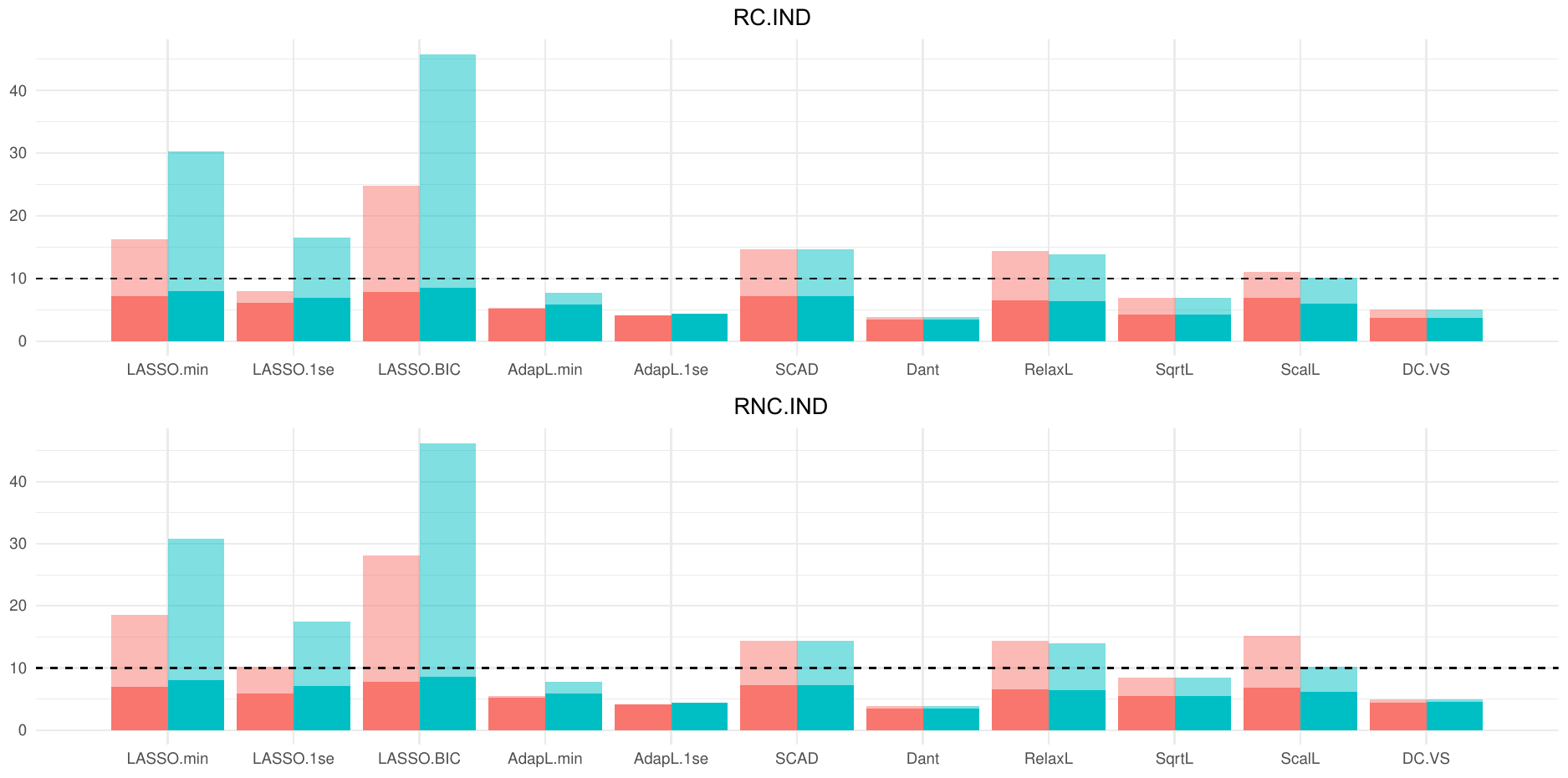}
	\caption{\label{cov_selected_comparison_LASSO_1b_1c_n_50}Number of important covariates (dark \textcolor{coralred}{left}/\textcolor{my_seagreen}{right} rectangular area) and noisy ones (soft \textcolor{lightcoral}{left}/\textcolor{blue-green}{right} rectangular area) for proposed algorithms taking $p=100$ and selected in terms of $\mathcolor{coralred}{\mathbf{X}^{\text{r}}}$/$\mathcolor{my_seagreen}{\mathbf{X}^{\text{us}}}$ in scenarios RC.IND (the first row) and RNC.IND (the second row) for $n=50$. The dashed line marks the $s=10$ value.}
\end{figure}

First, the RC.IND and RNC.IND scenarios are considered. Results taking $n=50$ are displayed in Figure \ref{cov_selected_comparison_LASSO_1b_1c_n_50} and Table \ref{summ_scenario_1_n_50}. All the studied procedures have a similar behavior in both scenarios, except for the LASSO approaches and the ScalL using $\mathbf{X}^{\text{r}}$: more noise is added in RNC.IND.

In Figure \ref{cov_selected_comparison_LASSO_1b_1c_n_50} we can appreciate as the use of $\mathbf{X}^{\text{r}}$ or $\mathbf{X}^{\text{us}}$ has an impact in the selection results in RC.IND and RNC.IND, except for the SCAD, Dant, SqrtL, and DC.VS algorithms. Using $\mathbf{X}^{\text{us}}$ seems to be helpful for recovering a few more elements of $S$, but paying the price of extra noise addition. Only for the ScalL happens the opposite. Regarding covariates selection, we can categorize algorithms into two types based on their selection strategy: the first group tries to recover the set $S$ completely (LASSO approaches, SCAD, RelaxL, SqrtL and ScalL), whereas the second one selects a representative subset of the relevant covariates (AdapL.min, AdapL.1se, Dant, and DC.VS). As a result, the first group recovers a greater amount of relevant terms, although more noise is added to the model. In these terms, the LASSO.BIC, and the LASSO.min are the noisiest algorithms. Conversely, the second one selects less variables, but all or almost all of these are important ones. In particular, only the AdapL.1se, Dant and DC.VS in RNC.IND methodologies avoid including noisy covariates. These three choose the relevant covariates with the highest scales a higher percentage of times than the ones with a value for the standard deviation less or equal to one. This fact is illustrated in Figure 20, collected in Section D.1 of the Appendix.

\begin{table}[htb] 
	\centering
	\adjustbox{max width=\textwidth}{
		\begin{tabular}{c cc cc cc cc}
			\toprule
			& \multicolumn{4}{c}{RC.IND} & \multicolumn{4}{c}{RNC.IND} \\
			\cmidrule(rl){2-5} \cmidrule(rl){6-9}
			& \multicolumn{2}{c}{\textbf{RAW}} &  \multicolumn{2}{c}{\textbf{UNIV.}}  & \multicolumn{2}{c}{\textbf{RAW}} &  \multicolumn{2}{c}{\textbf{UNIV.}} \\
			\cmidrule(r){2-3}  \cmidrule(rl){4-5} \cmidrule(rl){6-7} \cmidrule(l){8-9} 
			\vspace{-0.07cm}
			\rule{0pt}{0.35cm} \multirow{2}{*}{ \textbf{METHOD}} & {\footnotesize \textbf{MSE}} &  {\footnotesize \textbf{\% Dev}} & {\footnotesize \textbf{MSE}} &  {\footnotesize \textbf{\% Dev}} & {\footnotesize \textbf{MSE}} &  {\footnotesize \textbf{\% Dev}} & {\footnotesize \textbf{MSE}} &  {\footnotesize \textbf{\% Dev}} \\
			& {\scriptsize$(13.715)$} & {\scriptsize$(0.9)$} & {\scriptsize$(13.715)$} & {\scriptsize$(0.9)$} & {\scriptsize$(13.715)$} & {\scriptsize$(0.9)$} & {\scriptsize$(13.715)$} & {\scriptsize$(0.9)$} \\ 
			\cmidrule(r){1-3} \cmidrule(rl){4-5} \cmidrule(rl){6-7} \cmidrule(l){8-9} 
			\rule{0pt}{0.35cm} \textbf{LASSO.min} & 6.257 & 0.953 & 2.476 & 0.982 & 6.147 & 0.953 & 2.394 & 0.982  \rule[-0.25cm]{0pt}{0pt} \\ 
			\textbf{LASSO.1se} & \cellcolor[gray]{.9}12.624 & 0.905 & 7.092 & 0.947 & \cellcolor[gray]{.9}12.366 & 0.905 & 6.806 & 0.948 \rule[-0.25cm]{0pt}{0pt} \\ 
			\textbf{LASSO.BIC} & 2.468 & 0.983 & 0.024 & 1 & 1.860 & 0.987 & 0.022 & 1 \rule[-0.25cm]{0pt}{0pt} \\
			\textbf{AdapL.min} & \squaretext{17.46} & 0.87 & 11.604 & 0.914 & \squaretext{17.275} & 0.869 & 11.709 & 0.912  \rule[-0.25cm]{0pt}{0pt} \\    
			\textbf{AdapL.1se} & \squaretext{23.520} & 0.823 & \squaretext{21.975} & 0.834 & \squaretext{24.107} & 0.816 & \squaretext{22.031} & 0.832 \rule[-0.25cm]{0pt}{0pt} \\  
			\textbf{SCAD} & 5.892 & 0.956 & 5.892 & 0.956 & 6.181 & 0.952 & 6.181 & 0.952 \rule[-0.25cm]{0pt}{0pt} \\   
			\textbf{Dant} & \squaretext{30.202} & 0.775 & \squaretext{30.202} & 0.775 & \squaretext{30.584} & 0.769 & \squaretext{30.584} & 0.769 \rule[-0.25cm]{0pt}{0pt} \\  
			\textbf{RelaxL} & 9.298 & 0.929 & 9.789 & 0.925 & 9.250 & 0.929 & 9.849 & 0.924 \rule[-0.25cm]{0pt}{0pt} \\ 
			\textbf{SqrtL} & 9.789 & 0.925 & 9.789 & 0.925 & \cellcolor[gray]{.9}\squaretext{13.856} & 0.891 & \cellcolor[gray]{.9}\squaretext{13.856} & 0.891 \rule[-0.25cm]{0pt}{0pt} \\  
			\textbf{ScalL} & 8.186 & 0.938 & 11.054 & 0.914 & 6.950 & 0.947 & 11.038 & 0.914  \rule[-0.25cm]{0pt}{0pt} \\  
			\textbf{DC.VS} & 11.72 & 0.910 & 11.757 & 0.910 & 11.447 & 0.911 & \squaretext{14.233} & 0.886  \rule[-0.25cm]{0pt}{0pt} \\ 
			\bottomrule
		\end{tabular}
	}
	\caption[]{Comparison of all proposed algorithms for $p=100$ and $n=50$ using $\mathbf{X}^{\text{r}}$ and $\mathbf{X}^{\text{us}}$ in scenarios RC.IND and RNC.IND. Oracle values are in brackets. \hlc[gray!20]{Highlighted} terms are values in  \small{$[0.9\cdot\text{MSE},1.1\cdot\text{MSE}]$} (GI) and \dashbox[1.2cm][c]{squared} ones those that correct overestimation.}
	\label{summ_scenario_1_n_50}
\end{table}

Results about prediction for RC.IND and RNC.IND taking $n=50$ are collected in Table \ref{summ_scenario_1_n_50}. There, we can see that most procedures overestimate the prediction results and obtain less MSE and greater \%Dev than the oracle values. In this case, just LASSO.1se using $\mathbf{X}^{\text{r}}$ and SqrtL in RNC.IND obtain MSE values in the GI. Additionally, only AdapL.1se and Dant procedures can correct the overestimation in both frameworks. The AdapL.min using $\mathbf{X}^{\text{r}}$ and the SqrtL as well as the DC.VS using $\mathbf{X}^{\text{us}}$ in RNC.IND also correct this. The SqrtL procedure is the only one that verifies the two conditions, but just in the RNC.IND scenario.

\begin{figure}[h!]\centering
	\includegraphics[width=\linewidth]{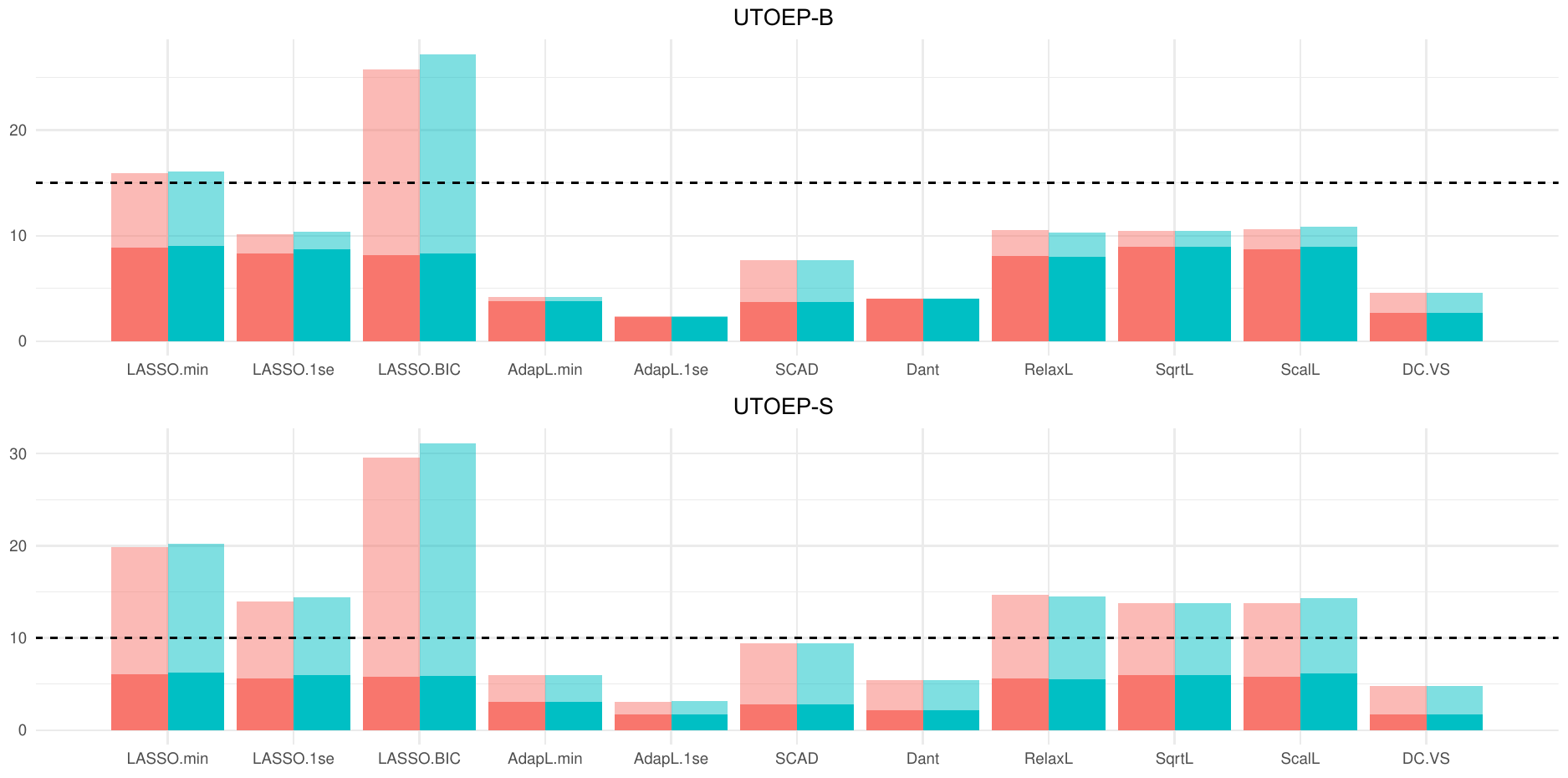}
	\caption{\label{cov_selected_comparison_LASSO_2a_2b_9_n_50}Number of important covariates (dark \textcolor{coralred}{left}/\textcolor{my_seagreen}{right} rectangular area) and noisy ones (soft \textcolor{lightcoral}{left}/\textcolor{blue-green}{right} rectangular area) for proposed algorithms taking $p=100$ and  $\mathcolor{coralred}{\mathbf{X}^{\text{r}}}$/$\mathcolor{my_seagreen}{\mathbf{X}^{\text{us}}}$ in scenarios UTOEP-B (the first row) and UTOEP-S (the second row) for $\rho=0.9$ and $n=50$. The dashed lines mark the $s=15$ and $s=10$ value for the first and the second row, respectively.}
\end{figure}

\begin{table}[h!] 
	\centering
	\small 
	\adjustbox{max width=\textwidth}{
		\begin{tabular}{c cc cc cc cc}
			\toprule
			& \multicolumn{4}{c}{UTOEP-B} & \multicolumn{4}{c}{UTOEP-S} \\
			\cmidrule(rl){2-5} \cmidrule(rl){6-9}
			& \multicolumn{2}{c}{\textbf{RAW}} &  \multicolumn{2}{c}{\textbf{UNIV.}} & \multicolumn{2}{c}{\textbf{RAW}} &  \multicolumn{2}{c}{\textbf{UNIV.}} \\
			\cmidrule(r){2-3}  \cmidrule(rl){4-5} \cmidrule(rl){6-7} \cmidrule(rl){8-9} 
			\vspace{-0.07cm}
			\rule{0pt}{0.35cm} \multirow{2}{*}{ \textbf{METHOD}} & {\footnotesize \textbf{MSE}} &  {\footnotesize \textbf{\% Dev}} & {\footnotesize \textbf{MSE}} &  {\footnotesize \textbf{\% Dev}} & {\footnotesize \textbf{MSE}} &  {\footnotesize \textbf{\% Dev}} & {\footnotesize \textbf{MSE}} &  {\footnotesize \textbf{\% Dev}} \\
			& {\scriptsize$(3.807)$} & {\scriptsize$(0.9)$} & {\scriptsize$(3.807)$} & {\scriptsize$(0.9)$} & {\scriptsize$(1.244)$} & {\scriptsize$(0.9)$} & {\scriptsize$(1.244)$} & {\scriptsize$(0.9)$} \\ 
			\cmidrule(r){1-3} \cmidrule(rl){4-5} \cmidrule(rl){6-7} \cmidrule(rl){8-9} 
			\rule{0pt}{0.35cm} \textbf{LASSO.min} & 1.914 & 0.948 & 1.898 & 0.948 & 0.514 & 0.955 & 0.513 & 0.955 \rule[-0.25cm]{0pt}{0pt} \\  
			\textbf{LASSO.1se} & 2.643 & 0.928 & 0.728 & 0.937 & 0.728 & 0.937 & 0.713 & 0.938 \rule[-0.25cm]{0pt}{0pt} \\  
			\textbf{LASSO.BIC} & 0.956 & 0.976 & 0.847 & 0.979 & 0.224 & 0.982 & 0.188 & 0.985 \rule[-0.25cm]{0pt}{0pt} \\  
			\textbf{AdapL.min} & 3.183 & 0.915 & 3.171 & 0.915 & 0.897 & 0.923 & 0.895 & 0.923 \rule[-0.25cm]{0pt}{0pt} \\   
			\textbf{AdapL.1se} & \squaretext{4.818} & 0.870 & \squaretext{4.713} & 0.872 & \squaretext{1.580} & 0.864 & \squaretext{1.554} & 0.866 \rule[-0.25cm]{0pt}{0pt} \\ 
			\textbf{SCAD} & 2.757 & 0.926 & 2.757 & 0.926 & \squaretext{1.554} & 0.866 & \squaretext{1.554} & 0.866 \rule[-0.25cm]{0pt}{0pt} \\ 
			\textbf{Dant} & \squaretext{4.82} & 0.87 & \squaretext{4.82} & 0.87 & \squaretext{1.554} & 0.866 & \squaretext{1.554} & 0.866 \rule[-0.25cm]{0pt}{0pt} \\  
			\textbf{RelaxL} & 2.671 & 0.928 & 2.714 & 0.926 & 0.729 & 0.937 & 0.734 & 0.936 \rule[-0.25cm]{0pt}{0pt} \\ 
			\textbf{SqrtL} & 2.623 & 0.929 & 2.623 & 0.929 & 0.734 & 0.936 & 0.749 & 0.935 \rule[-0.25cm]{0pt}{0pt} \\  
			\textbf{ScalL} & 2.562 & 0.930 & 2.534 & 0.931 & 0.743 & 0.935 & 0.726 & 0.937 \rule[-0.25cm]{0pt}{0pt} \\ 
			\textbf{DC.VS} & \cellcolor[gray]{.9}\squaretext{3.958} & 0.894 & \cellcolor[gray]{.9}\squaretext{3.958} & 0.894 & \squaretext{1.402} & 0.880 & \squaretext{1.402} & 0.880 \rule[-0.25cm]{0pt}{0pt} \\  
			\bottomrule
		\end{tabular}
	}
	\caption[]{Comparison of all proposed algorithms for $p=100$, $n=50$ and $\rho=0.9$ using $\mathbf{X}^{\text{r}}$ and $\mathbf{X}^{\text{us}}$ in UTOEP scenario. Oracle values are in brackets. \hlc[gray!20]{Highlighted} terms are values in  \small{$[0.9\cdot\text{MSE},1.1\cdot\text{MSE}]$} (GI) and \dashbox[1.2cm][c]{squared} ones those that correct overestimation. }
	\label{summ_scenario_2_n_50}
\end{table}

Next, results for competitors in UTOEP scenarios (UTOEP-B and UTOEP-S) are analyzed. Results taking $\rho=0.9$ are summarized in Figure \ref{cov_selected_comparison_LASSO_2a_2b_9_n_50} and Table \ref{summ_scenario_2_n_50}. Those for the $\rho=0.5$ case are displayed in Figure 26 and Table 19 in Section D.2 of the Appendix. In all cases, one can observe a similar behavior for all considered procedures using $\mathbf{X}^{\text{r}}$ or $\mathbf{X}^{\text{us}}$ approaches. In these settings, a distinction in terms of the selection philosophy can also be made: approaches searching for a complete recovery of the $s$ relevant variables (LASSO methodologies, AdapL.min, SCAD, RelaxL, SqrtL, ScalL and DC.VS) and those that employ the dependence structure and select a representative subset of these (AdapL.1se and Dant). This last is especially remarkable in the $\rho=0.9$ case (see Figure \ref{cov_selected_comparison_LASSO_2a_2b_9_n_50}). In particular, for $\rho=0.9$, one can also add the AdapL.min and DC.VS to the second group. In Table 14 of Section B in the Appendix, it can be seen that with a bunch of covariates smaller than $s$, it is possible to explain a large amount of variability. As a result, this second methodology also makes sense in the UTOEP scenarios. As can be seen in Figure \ref{cov_selected_comparison_LASSO_2a_2b_9_n_50} for $\rho=0.9$, and in Figure 26 in Section D.2 of the Appendix for $\rho=0.5$, algorithms searching for the complete recovery of $S$ also select more noise than the other ones. In contrast, the procedures that make use of the dependence structure tend to select just relevant terms. Hence, the first group increases the proportion of true positives, although more noisy covariates are added, whereas the second group decreases the rate of false discoveries. This last can be appreciated seeing the percentage of times each of the relevant covariates and high correlated close ones enters the model in UTOEP-B and UTOEP-S for AdapL.1se, Dant, and DC.VS. Results for UTOEP-B and UTOEP-S are displayed in Figure 24 and Figure 35, respectively, in Section D.2 of the Appendix. Concerning UTOEP-B results, one can see as these three procedures select the first $s=15$ relevant covariates the highest percentage of times. Only the DC.VS selects some noisy covariates in the neighborhood of the 15$^{th}$ term for $\rho=0.9$. Related to UTOEP-S, these approaches perform well selecting the important variables with high probability. For $\rho=0.5$, the remaining covariates are selected a negligible percentage of times, especially for AdapL.1se and Dant. In contrast, for the $\rho=0.9$ case, the confusion phenomenon intensifies. This translates in a higher selection of noisy covariates related to relevant ones, especially for Dant and DC.VS algorithms.

Concerning prediction, we note that all algorithms overestimate the results, but for the AdapL.1se, SCAD in UTOEP-S with $\rho=0.9$, Dant, SqrtL in UTOEP-S with $\rho=0.5$ and DC.VS (see Table \ref{summ_scenario_2_n_50}). These are the procedures that search for a representative subset of $S$ in the selection procedure. Only the ScalL in UTOEP-B with $\rho=0.5$ and DC.VS in UTOEP-B with $\rho=0.9$ verify that its associated MSE is in the GI. Indeed, as expected, no distinction is appreciated between $\mathbf{X}^{\text{r}}$ or $\mathbf{X}^{\text{us}}$ implementation.

\begin{figure}[h!]\centering
	\includegraphics[width=\linewidth]{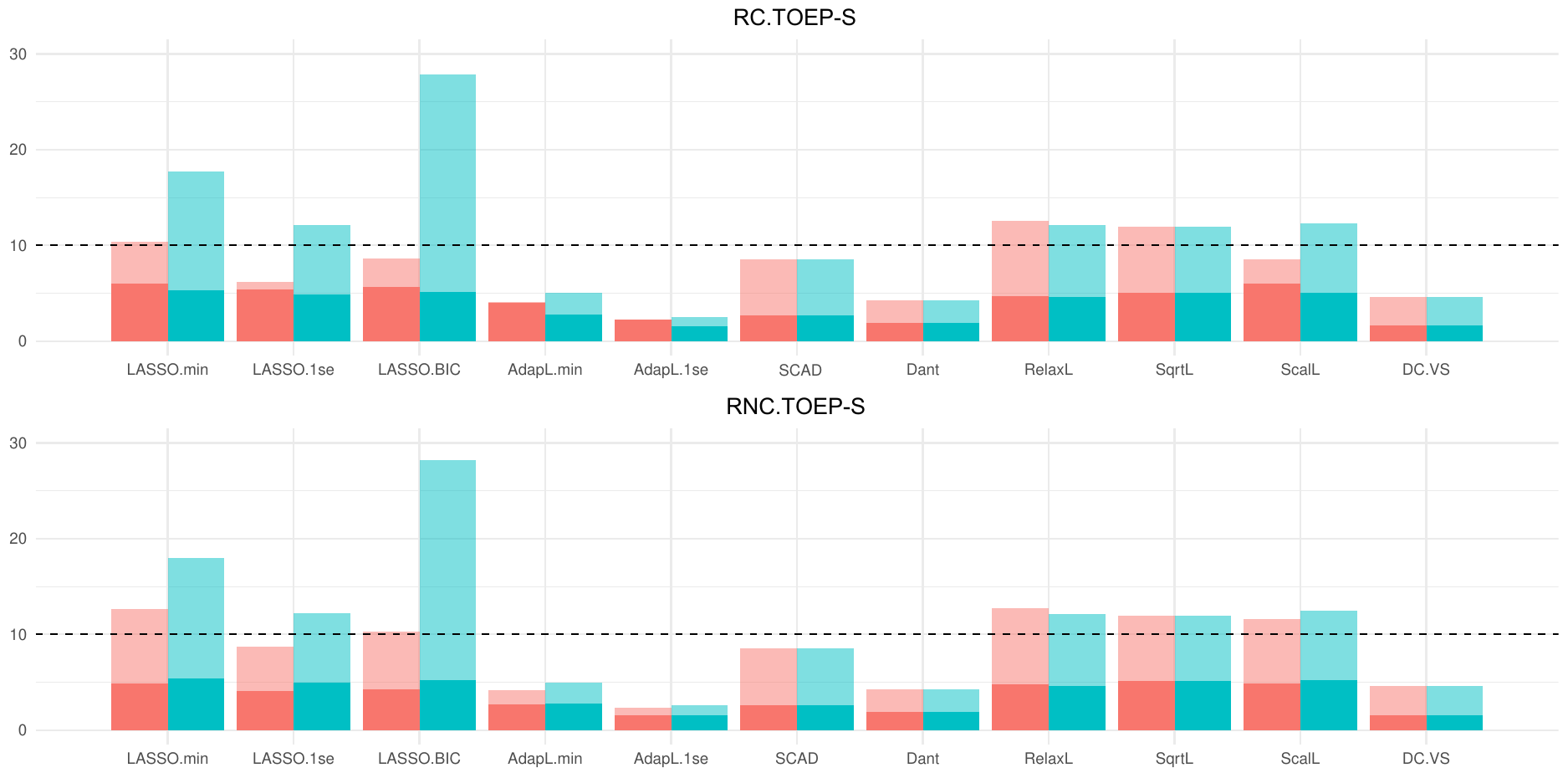}
	\caption{\label{cov_selected_comparison_LASSO_3a_3b_rho_0.9_n_50}Number of important covariates (dark \textcolor{coralred}{left}/\textcolor{my_seagreen}{right} rectangular area) and noisy ones (soft \textcolor{lightcoral}{left}/\textcolor{blue-green}{right} rectangular area) for proposed algorithms taking $p=100$ and selected in terms of $\mathcolor{coralred}{\mathbf{X}^{\text{r}}}$/$\mathcolor{my_seagreen}{\mathbf{X}^{\text{us}}}$ in scenarios RC.TOEP-S (the first row) and RNC.TOEP-S (the second row) for $\rho=0.9$ and $n=50$. The dashed line marks the $s=10$ value.}
\end{figure}

\begin{table}[h!] 
	\centering
	\small 
	\adjustbox{max width=\textwidth}{
		\begin{tabular}{c cc cc cc cc}
			\toprule
			& \multicolumn{4}{c}{RC.TOEP-S} & \multicolumn{4}{c}{RNC.TOEP-S} \\
			\cmidrule(rl){2-5} \cmidrule(rl){6-9}
			& \multicolumn{2}{c}{\textbf{RAW}} &  \multicolumn{2}{c}{\textbf{UNIV.}} & \multicolumn{2}{c}{\textbf{RAW}} &  \multicolumn{2}{c}{\textbf{UNIV.}} \\
			\cmidrule(r){2-3}  \cmidrule(rl){4-5} \cmidrule(rl){6-7} \cmidrule(rl){8-9} 
			\vspace{-0.07cm}
			\rule{0pt}{0.35cm} \multirow{2}{*}{ \textbf{METHOD}} & {\footnotesize \textbf{MSE}} &  {\footnotesize \textbf{\% Dev}} & {\footnotesize \textbf{MSE}} &  {\footnotesize \textbf{\% Dev}} & {\footnotesize \textbf{MSE}} &  {\footnotesize \textbf{\% Dev}} & {\footnotesize \textbf{MSE}} &  {\footnotesize \textbf{\% Dev}} \\
			& {\scriptsize$(7.818)$} & {\scriptsize$(0.9)$} & {\scriptsize$(7.818)$} & {\scriptsize$(0.9)$} & {\scriptsize$(7.818)$} & {\scriptsize$(0.9)$} & {\scriptsize$(7.818)$} & {\scriptsize$(0.9)$} \\ 
			\cmidrule(r){1-3} \cmidrule(rl){4-5} \cmidrule(rl){6-7} \cmidrule(rl){8-9} 
			\rule{0pt}{0.35cm} \textbf{LASSO.min} & 5.060 & 0.932 & 3.577 & 0.952 & 4.828 & 0.935 & 3.570 & 0.952 \rule[-0.25cm]{0pt}{0pt} \\ 
			\textbf{LASSO.1se} & 6.927 & 0.907 & 4.871 & 0.934 & 6.500 & 0.913 & 4.856 & 0.935 \rule[-0.25cm]{0pt}{0pt} \\ 
			\textbf{LASSO.BIC} & 5.258 & 0.929 & 1.596 & 0.980 & 5.265 & 0.930 & 1.505 & 0.981 \rule[-0.25cm]{0pt}{0pt} \\ 
			\textbf{AdapL.min} & \cellcolor[gray]{.9}\squaretext{8.358} & 0.888 & 6.077 & 0.919 & \cellcolor[gray]{.9}7.683 & 0.898 & 6.157 & 0.918 \rule[-0.25cm]{0pt}{0pt} \\    
			\textbf{AdapL.1se} & \squaretext{12.461} & 0.83 & \squaretext{10.416} & 0.860 & \squaretext{11.898} & 0.841 & \squaretext{10.349} & 0.862 \rule[-0.25cm]{0pt}{0pt} \\  
			\textbf{SCAD} & 5.348 & 0.928 & 5.348 & 0.928 & \squaretext{10.349} & 0.862 & \squaretext{10.349} & 0.862 \rule[-0.25cm]{0pt}{0pt} \\  
			\textbf{Dant} & \squaretext{10.899} & 0.853 & \squaretext{10.899} & 0.853 & \squaretext{10.905} & 0.855 & \squaretext{10.905} & 0.855 \rule[-0.25cm]{0pt}{0pt} \\  
			\textbf{RelaxL} & 4.925 & 0.933 & 5.051 & 0.932 & 4.899 & 0.934 & 5.083 & 0.932 \rule[-0.25cm]{0pt}{0pt} \\   
			\textbf{SqrtL} & 5.051 & 0.932 & 5.051 & 0.932 & 4.971 & 0.933 & 4.971 & 0.933 \rule[-0.25cm]{0pt}{0pt} \\  
			\textbf{ScalL} & 5.562 & 0.925 & 4.843 & 0.934 & 5.114 & 0.931 & 4.838 & 0.935 \rule[-0.25cm]{0pt}{0pt} \\
			\textbf{DC.VS} & \squaretext{8.724} & 0.883 & \squaretext{8.724} & 0.883 & \squaretext{8.649} & 0.885 & \squaretext{8.649} & 0.885 \rule[-0.25cm]{0pt}{0pt} \\  
			\bottomrule
		\end{tabular}
	}
	\caption[]{Comparison of all proposed algorithms for $p=100$, $n=50$ and $\rho=0.9$ using $\mathbf{X}^{\text{r}}$ and $\mathbf{X}^{\text{us}}$ in Scenario 3. Oracle values are in brackets. \hlc[gray!20]{Highlighted} terms are values in  \small{$[0.9\cdot\text{MSE},1.1\cdot\text{MSE}]$} (GI) and \dashbox[1.2cm][c]{squared} ones those that correct overestimation. }
	\label{summ_scenario_3_n_50}
\end{table}

Eventually, we test the performance of the proposed algorithms in the TOEP-S framework (RC.TOEP-S and RNC.TOEP-S). Results for $\rho=0.9$ are summarized in Figure \ref{cov_selected_comparison_LASSO_3a_3b_rho_0.9_n_50} and Table \ref{summ_scenario_3_n_50}, while those for $\rho=0.5$ are displayed in Figure 34 and Table 23 of Section D.3 in the Appendix. In view of the $\mathbf{X}^{\text{r}}$ and $\mathbf{X}^{\text{us}}$ results, some algorithms keep the same selection strategy in both cases (SCAD, Dant, SqrtL, and DC.VS). In contrast, the use of $\mathbf{X}^{\text{us}}$ in the remaining procedures increases the number of selected terms, adding more noise in this process, except for RelaxL. A similar behavior is observed for the $\rho=0.5$ case adding more noise in this last (see Figure 34 of Section D.3 in the Appendix). In RC.TOEP-S with $\rho=0.9$, the use of $\mathbf{X}^{\text{us}}$ does not improve the selection results: this selects a similar or least number of relevant terms but adds quite much noise. This framework exemplifies how univariate standardization can disrupt the dependence structure and worsen the selection results, even having covariates in different scales. In contrast, in RNC.TOEP-S for both $\rho$ values, the $\mathbf{X}^{\text{us}}$ approach selects some more relevant terms but pays a high price in terms of extra noise addition. Again, there are algorithms making use of the dependence structure and selecting a representative subset of the $s$ relevant terms (AdapL.min, AdapL.1se, Dant and DC.VS), whereas others search for the complete recovery of $S$. This strategy makes sense if we notice that an efficient number of covariates needed to explain a great percentage of variability is less than $s$ (see Table 15 in Section B of the Appendix). Furthermore, the algorithms in this first group select the relevant covariates with the greatest scales ($j=15$, $18$, $21$, $24$, $27$, $30$) a high percentage of times for both scenarios and $\rho=0.5$ (see Figure 35 in Section D.3 of the Appendix) as well as $\rho=0.9$ (see Figure 36 in Section D.3 of the Appendix). However, in the $\rho=0.9$ case, some important terms are interchanged with quite correlated to the previous ones because of the strong dependence structure.

In Table \ref{summ_scenario_3_n_50}, results for prediction in RC.TOEP-S and RNC.TOEP-S taking $n=50$ and $\rho=0.9$ are displayed. Similar values are appreciated for $\mathbf{X}^{\text{r}}$ and $\mathbf{X}^{\text{us}}$, except for those algorithms where the standardization has an impact on the selection procedure. All procedures tend to overestimate the results, but for some related to the representative subset strategy (AdapL.1se, Dant and DC.VS). We can also include the SCAD procedure in the RNC.TOEP-S scenario. In this case, only the AdapL.min verifies that its prediction error is between the $[0.9\cdot\text{MSE},1.1\cdot\text{MSE}]$ values using $\mathbf{X}^{\text{r}}$.

\subsection{Context of $\bf n\geq p$}

\begin{figure}[htb]\centering
	\includegraphics[width=\linewidth]{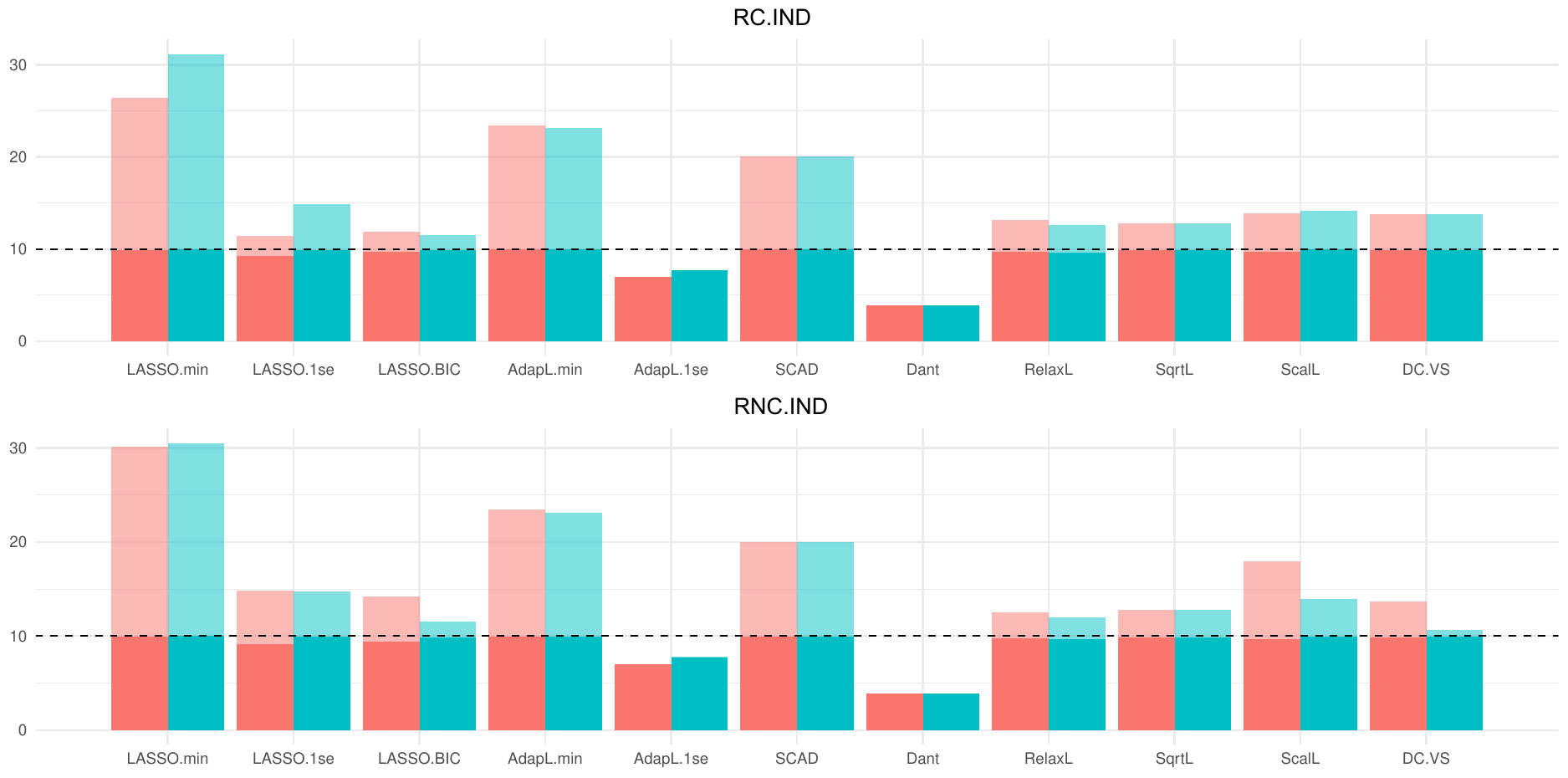}
	\caption{\label{cov_selected_comparison_LASSO_1b_1c}Number of important covariates (dark \textcolor{coralred}{left}/\textcolor{my_seagreen}{right} rectangular area) and noisy ones (soft \textcolor{lightcoral}{left}/\textcolor{blue-green}{right} rectangular area) for proposed algorithms taking $p=100$ and selected in terms of $\mathcolor{coralred}{\mathbf{X}^{\text{r}}}$/$\mathcolor{my_seagreen}{\mathbf{X}^{\text{us}}}$ in scenarios RC.IND (the first row) and RNC.IND (the second row) for $n=300$. The dashed line marks the $s=10$ value.}
\end{figure}

Results for RC.IND and RNC.IND scenarios are collected in Figure \ref{cov_selected_comparison_LASSO_1b_1c} and Table \ref{summ_scenario_1} taking $n=300$. In this $n>p$ case, almost all algorithms have a similar behavior between $\mathbf{X}^{\text{r}}$ and $\mathbf{X}^{\text{us}}$ approaches. Exceptions are the LASSO.min as well as LASSO.1se using $\mathbf{X}^{\text{us}}$ in RC.IND, and the LASSO.BIC, ScalL and DC.VS with $\mathbf{X}^{\text{r}}$ in RNC.IND. All of these add more noisy covariates than their standardization counterparts. Similar to the $p>n$ framework, one can distinguish between algorithms making use of the different scale structures selecting a subset of $S$ (AdapL.1se and Dant) and those tending to select all the $s$ relevant terms (remaining ones). In particular, AdapL.1se and Dant procedures recover the relevant covariates with the largest scales. An example of this fact is displayed in Figure 21, collected in Section D.1 of the Appendix. Furthermore, these algorithms are the only ones able to guarantee the absence of noise in the selection process. Now, the algorithms that search for a fully recovery of $S$ are able to select all the relevant terms adding less noise to the model in comparison to the $p>n$ framework (see Figure \ref{cov_selected_comparison_LASSO_1b_1c_n_50}).

Concerning prediction in RC.IND and RNC.IND scenarios, results are collected in Table \ref{summ_scenario_1}. There, one can see that almost all procedures overestimate the prediction results, MSE or \%Dev values less and greater than the oracle ones, respectively; but for the AdapL.1se and the Dant. These two algorithms are the ones that select a subset of $S$ in the covariates selection procedure. In contrast, only the LASSO.1se using $\mathbf{X}^{\text{r}}$, LASSO.BIC, AdapL.1se using $\mathbf{X}^{\text{us}}$, RelaxL, SqrtL and DC.VS verify that their estimations of the MSE are in the GI.

\begin{table}[htb] 
	\centering
	\adjustbox{max width=\textwidth}{
		\begin{tabular}{c cc cc cc cc}
			\toprule
			& \multicolumn{4}{c}{RC.IND} & \multicolumn{4}{c}{RNC.IND} \\
			\cmidrule(rl){2-5} \cmidrule(rl){6-9}
			& \multicolumn{2}{c}{\textbf{RAW}} &  \multicolumn{2}{c}{\textbf{UNIV.}}  & \multicolumn{2}{c}{\textbf{RAW}} &  \multicolumn{2}{c}{\textbf{UNIV.}} \\
			\cmidrule(r){2-3}  \cmidrule(rl){4-5} \cmidrule(rl){6-7} \cmidrule(l){8-9} 
			\vspace{-0.07cm}
			\rule{0pt}{0.35cm} \multirow{2}{*}{ \textbf{METHOD}} & {\footnotesize \textbf{MSE}} &  {\footnotesize \textbf{\% Dev}} & {\footnotesize \textbf{MSE}} &  {\footnotesize \textbf{\% Dev}} & {\footnotesize \textbf{MSE}} &  {\footnotesize \textbf{\% Dev}} & {\footnotesize \textbf{MSE}} &  {\footnotesize \textbf{\% Dev}} \\
			& {\scriptsize$(13.715)$} & {\scriptsize$(0.9)$} & {\scriptsize$(13.715)$} & {\scriptsize$(0.9)$} & {\scriptsize$(13.715)$} & {\scriptsize$(0.9)$} & {\scriptsize$(13.715)$} & {\scriptsize$(0.9)$} \\ 
			\cmidrule(r){1-3} \cmidrule(rl){4-5} \cmidrule(rl){6-7} \cmidrule(l){8-9} 
			\rule{0pt}{0.35cm} \textbf{LASSO.min} & 10.972 & 0.919 & 10.638 & 0.922 & 10.866 & 0.920 & 10.677 & 0.921 \rule[-0.25cm]{0pt}{0pt} \\
			\textbf{LASSO.1se} & \cellcolor[gray]{.9}12.972 & 0.904 & 12.174 & 0.910 & \cellcolor[gray]{.9}12.813 & 0.891 & 12.180 & 0.891 \rule[-0.25cm]{0pt}{0pt} \\ 
			\textbf{LASSO.BIC} & \cellcolor[gray]{.9}12.671 & 0.907 & \cellcolor[gray]{.9}12.739 & 0.906 & \cellcolor[gray]{.9}12.729 & 0.906 & \cellcolor[gray]{.9}12.722 & 0.906 \rule[-0.25cm]{0pt}{0pt} \\   
			\textbf{AdapL.min} & 11.115 & 0.918 & 11.138 & 0.918 & 11.117 & 0.918 & 11.136 & 0.918 \rule[-0.25cm]{0pt}{0pt} \\  
			\textbf{AdapL.1se} & \squaretext{15.921} & 0.883 & \cellcolor[gray]{.9}\squaretext{15.024} & 0.889 & \squaretext{15.901} & 0.883 & \cellcolor[gray]{.9}\squaretext{14.977} & 0.890 \rule[-0.25cm]{0pt}{0pt} \\ 
			\textbf{SCAD} & 11.441 & 0.916 & 11.441 & 0.916 & 11.453 & 0.915 & 11.453 & 0.915  \rule[-0.25cm]{0pt}{0pt} \\  
			\textbf{Dant} & \squaretext{29.407} & 0.784 & \squaretext{29.407} & 0.784 & \squaretext{29.549} & 0.782 & \squaretext{29.549} & 0.782  \rule[-0.25cm]{0pt}{0pt} \\  
			\textbf{RelaxL} & \cellcolor[gray]{.9}12.630 & 0.907 & \cellcolor[gray]{.9}12.754 & 0.906 & \cellcolor[gray]{.9}12.687 & 0.906 & \cellcolor[gray]{.9}12.813 & 0.905 \rule[-0.25cm]{0pt}{0pt} \\ 
			\textbf{SqrtL} & \cellcolor[gray]{.9}12.521 & 0.908 & \cellcolor[gray]{.9}12.521 & 0.908 & \cellcolor[gray]{.9}12.508 & 0.908 & \cellcolor[gray]{.9}12.508 & 0.908  \rule[-0.25cm]{0pt}{0pt} \\  
			\textbf{ScalL} & 12.324 & 0.909 & 12.270 & 0.910 & 12.096 & 0.911 & 12.286 & 0.909  \rule[-0.25cm]{0pt}{0pt} \\  
			\textbf{DC.VS} & \cellcolor[gray]{.9}12.353 & 0.909 & \cellcolor[gray]{.9}12.353 & 0.909 & \cellcolor[gray]{.9}12.371 & 0.909 & \cellcolor[gray]{.9}12.748 & 0.900 \rule[-0.25cm]{0pt}{0pt} \\ 
			\bottomrule
		\end{tabular}  
	}
	\caption[]{Comparison of all proposed algorithms for $p=100$ and $n=300$ using $\mathbf{X}^{\text{r}}$ and $\mathbf{X}^{\text{us}}$ in scenarios RC.IND and RNC.IND. Oracle values are in brackets. \hlc[gray!20]{Highlighted} terms are values in  \small{$[0.9\cdot\text{MSE},1.1\cdot\text{MSE}]$} (GI) and \dashbox[1.2cm][c]{squared} ones those that correct overestimation.. }
	\label{summ_scenario_1}
\end{table}

\begin{figure}[htb]\centering
	\includegraphics[width=\linewidth]{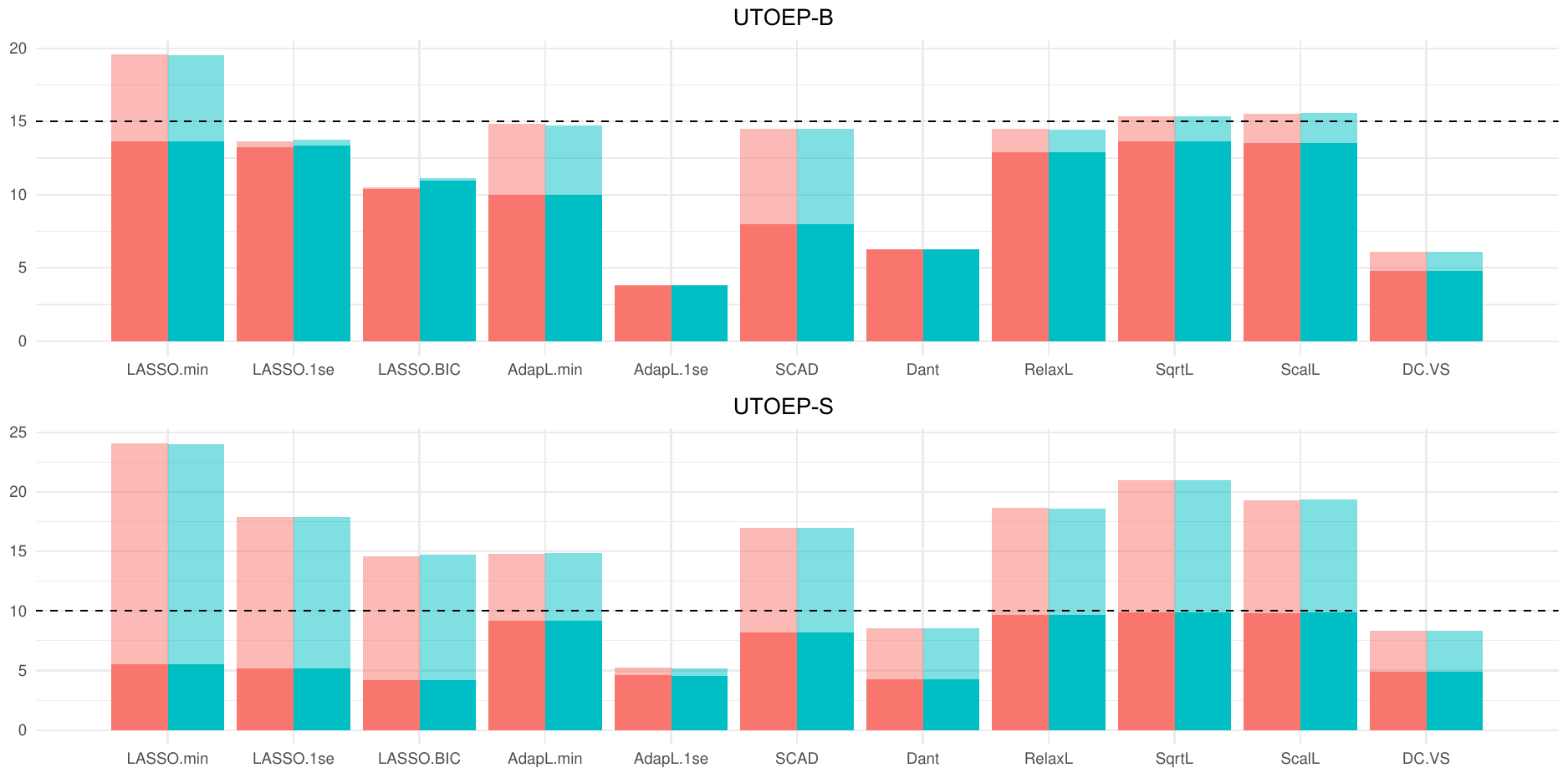}
	\caption{\label{cov_selected_comparison_LASSO_2a_2b_9}Number of important covariates (dark \textcolor{coralred}{left}/\textcolor{my_seagreen}{right} rectangular area) and noisy ones (soft \textcolor{lightcoral}{left}/\textcolor{blue-green}{right} rectangular area) for proposed algorithms taking $p=100$ and  $\mathcolor{coralred}{\mathbf{X}^{\text{r}}}$/$\mathcolor{my_seagreen}{\mathbf{X}^{\text{us}}}$ in scenarios UTOEP-B (the first row) and UTOEP-S (the second row) for $\rho=0.9$ and $n=300$. The dashed lines mark the $s=15$ and $s=10$ value for the first and the second row, respectively.}
\end{figure}

Next, we analyze the results in the UTOEP scenarios for $p>n$. Results for $n=300$ and taking $\rho=0.9$ are displayed in Figure \ref{cov_selected_comparison_LASSO_2a_2b_9} and Table \ref{summ_scenario_2}. Results for $\rho=0.5$ are collected in Figure 26 and Table 20 in Section D.2 of the Appendix. In both UTOEP-B as well as UTOEP-S, one can see that the results are pretty similar in all studied procedures for $\mathbf{X}^{\text{r}}$ or $\mathbf{X}^{\text{us}}$, indistinctly. As noticed in the $p>n$ framework (see Figure \ref{cov_selected_comparison_LASSO_2a_2b_9_n_50}), this was an expected behavior, since all covariates are in unitary scales. In this $n>p$ framework, we can see differences in the way of proceeding for the considered algorithms. In the UTOEP-B case, only the AdapL.1se, Dant and DC.VS taking $\rho=0.9$ employ the dependence structure and select a representative subset of the $s=15$ relevant variables without no or little noise inclusion. As previously mentioned, a number of relevant covariates less than $s=15$ are enough to explain a great percentage of the data variability in this scenario (see Table 14 of Section B in the Appendix). This is especially remarkable for $\rho=0.9$. As a result, this selection strategy seems a proper one. Furthermore, we can see that these three procedures, AdapL.1se, Dant and DC.VS, help to decrease the number of false negatives, selecting the first $s=15$ variables with the highest probability and no noisy terms, except for the DC.VS (see Figure 27 in Section D.2 of the Appendix). In contrast, the rest of the algorithms tend to recover all relevant terms, adding noisy covariates as a trade-off. This last translates into an increment in the proportion of true positives. However, in the case of strong dependence taking $\rho=0.9$ (see Figure \ref{cov_selected_comparison_LASSO_2a_2b_9}), no procedure is able to achieve a complete recovery of $S$. Instead, some important covariates are exchanged for irrelevant ones that are pretty correlated. On the other hand, now in the UTOEP-S scenario with $n>p$, all algorithms search for all $s=10$ relevant variables in the $\rho=0.5$ case (see Figure 26 in Section D.2 of the Appendix). Still, only some of these make use of the dependence structure selecting less than $s=10$ covariates in the $\rho=0.9$ framework (see Figure \ref{cov_selected_comparison_LASSO_2a_2b_9}). As a reason for this behavior, one can see that with less than $s=10$ relevant terms, it is possible to explain a high amount of variability in the UTOEP-S scenario, especially in the $\rho=0.9$ case (see Table 14 of Section B in the Appendix). In particular, only the AdapL.1se, Dant and DC.VS are the ones that make use of the strong dependence structure for $\rho=0.9$ selecting the relevant terms, or strongly correlated ones, the highest percentage of times (see Figure 28 in Section D.2 of the Appendix). These approaches are also some of the algorithms that totally recover $S$ in the $\rho=0.5$ case with small noise addition.

\begin{table}[h!] 
	\centering
	\small 
	\adjustbox{max width=\textwidth}{
		\begin{tabular}{c cc cc cc cc}
			\toprule
			& \multicolumn{4}{c}{UTOEP-B} & \multicolumn{4}{c}{UTOEP-S} \\
			\cmidrule(rl){2-5} \cmidrule(rl){6-9}
			& \multicolumn{2}{c}{\textbf{RAW}} &  \multicolumn{2}{c}{\textbf{UNIV.}} & \multicolumn{2}{c}{\textbf{RAW}} &  \multicolumn{2}{c}{\textbf{UNIV.}} \\
			\cmidrule(r){2-3}  \cmidrule(rl){4-5} \cmidrule(rl){6-7} \cmidrule(rl){8-9} 
			\vspace{-0.07cm}
			\rule{0pt}{0.35cm} \multirow{2}{*}{ \textbf{METHOD}} & {\footnotesize \textbf{MSE}} &  {\footnotesize \textbf{\% Dev}} & {\footnotesize \textbf{MSE}} &  {\footnotesize \textbf{\% Dev}} & {\footnotesize \textbf{MSE}} &  {\footnotesize \textbf{\% Dev}} & {\footnotesize \textbf{MSE}} &  {\footnotesize \textbf{\% Dev}} \\
			& {\scriptsize$(3.807)$} & {\scriptsize$(0.9)$} & {\scriptsize$(3.807)$} & {\scriptsize$(0.9)$} & {\scriptsize$(1.244)$} & {\scriptsize$(0.9)$} & {\scriptsize$(1.244)$} & {\scriptsize$(0.9)$} \\ 
			\cmidrule(r){1-3} \cmidrule(rl){4-5} \cmidrule(rl){6-7} \cmidrule(rl){8-9} 
			\rule{0pt}{0.35cm} \textbf{LASSO.min} &  \cellcolor[gray]{.9}3.525 & 0.910 &  \cellcolor[gray]{.9}3.526 & 0.910 & 1.096 & 0.911 & 1.098 & 0.911 \rule[-0.25cm]{0pt}{0pt} \\ 
			\textbf{LASSO.1se} &  \cellcolor[gray]{.9}3.715 & 0.905 &  \cellcolor[gray]{.9}3.715 & 0.905 &  \cellcolor[gray]{.9}1.154 & 0.906 &  \cellcolor[gray]{.9}1.154 & 0.906 \rule[-0.25cm]{0pt}{0pt} \\ 
			\textbf{LASSO.BIC} &  \cellcolor[gray]{.9}\squaretext{3.822} & 0.902 &  \cellcolor[gray]{.9}3.800 & 0.903 &  \cellcolor[gray]{.9}1.183 & 0.904 &  \cellcolor[gray]{.9}1.180 & 0.904 \rule[-0.25cm]{0pt}{0pt} \\ 
			\textbf{AdapL.min} &  \cellcolor[gray]{.9}3.513 & 0.910 &  \cellcolor[gray]{.9}3.518 & 0.910 & 1.117 & 0.909 & 1.116 & 0.909 \rule[-0.25cm]{0pt}{0pt} \\  
			\textbf{AdapL.1se} & \squaretext{4.540} & 0.884 & \squaretext{4.533} & 0.884 & \squaretext{1.501} & 0.878 & \squaretext{1.509} & 0.877 \rule[-0.25cm]{0pt}{0pt} \\
			\textbf{SCAD} &  \cellcolor[gray]{.9}3.602 & 0.908 &  \cellcolor[gray]{.9}3.602 & 0.908 & 1.116 & 0.909 & 1.116 & 0.909 \rule[-0.25cm]{0pt}{0pt} \\
			\textbf{Dant} & \squaretext{4.886} & 0.875 & \squaretext{4.886} & 0.875 & \squaretext{1.700} & 0.862 & \squaretext{1.700} & 0.862 \rule[-0.25cm]{0pt}{0pt} \\ 
			\textbf{RelaxL} &  \cellcolor[gray]{.9}3.683 & 0.906 &  \cellcolor[gray]{.9}3.685 & 0.906 &  \cellcolor[gray]{.9}1.143 & 0.907 &  \cellcolor[gray]{.9}1.144 & 0.907 \rule[-0.25cm]{0pt}{0pt} \\
			\textbf{SqrtL} &  \cellcolor[gray]{.9}3.654 & 0.906 &  \cellcolor[gray]{.9}3.654 & 0.906 &  \cellcolor[gray]{.9}1.139 & 0.908 &  \cellcolor[gray]{.9}1.139 & 0.908 \rule[-0.25cm]{0pt}{0pt} \\ 
			\textbf{ScalL} &  \cellcolor[gray]{.9}3.636 & 0.907 &  \cellcolor[gray]{.9}3.634 & 0.907 &  \cellcolor[gray]{.9}1.137 & 0.908 &  \cellcolor[gray]{.9}1.137 & 0.908 \rule[-0.25cm]{0pt}{0pt} \\
			\textbf{DC.VS} & \squaretext{4.194} & 0.892 & \squaretext{4.194} & 0.892 &  \cellcolor[gray]{.9}\squaretext{1.347} & 0.891 &  \cellcolor[gray]{.9}\squaretext{1.347} & 0.891 \rule[-0.25cm]{0pt}{0pt} \\ 
			\bottomrule
		\end{tabular}
	}
	\caption[]{Comparison of all proposed algorithms for $p=100$, $n=300$ and $\rho=0.9$ using $\mathbf{X}^{\text{r}}$ and $\mathbf{X}^{\text{us}}$ in UTOEP scenario. Oracle values are in brackets. \hlc[gray!20]{Highlighted} terms are values in  \small{$[0.9\cdot\text{MSE},1.1\cdot\text{MSE}]$} (GI) and \dashbox[1.2cm][c]{squared} ones those that correct overestimation.. }
	\label{summ_scenario_2}
\end{table}

Prediction results for UTOEP-B and UTOEP-S taking $\rho=0.9$ are summarized in Table \ref{summ_scenario_2}. Results for $\rho=0.5$ can be checked in Table 20 in Section D.2 of the Appendix. Once again, one can see as the results for MSE and \%Dev tend to be overestimated in all cases. For UTOEP-B, only the AdapL.1se, Dant and DC.VS taking $\rho=0.9$ are able to correct this overestimation. The remaining procedures overestimate the results, but verify that their associated MSE are in the GI in the $\rho=0.9$ case. Only the LASSO.1se, LASSO.BIC and RelaxL also verify this condition in the $\rho=0.5$ framework. Related to UTOEP-S, only the Dant procedure corrects the overestimation for $\rho=0.5$ and $\rho=0.9$, while AdapL.1se and DC.VS only in the $\rho=0.9$ case. The LASSO.BIC, AdapL.1se, SCAD, Dant, RelaxL, SqrtL and DC.VS verifies that their associated MSE value is in the GI for the $\rho=0.5$ scenario (see Table 20 in Section D.2 of the Appendix). In contrast this happens for the LASSO.1se, LASSO.BIC, RelaxL, SqrtL, ScalL and DC.VS in the $\rho=0.9$ case (see Table \ref{summ_scenario_2}). Similar prediction results are obtained for the different configurations using $\mathbf{X}^{\text{r}}$ and $\mathbf{X}^{\text{us}}$.

\begin{figure}[htb]\centering
	\includegraphics[width=\linewidth]{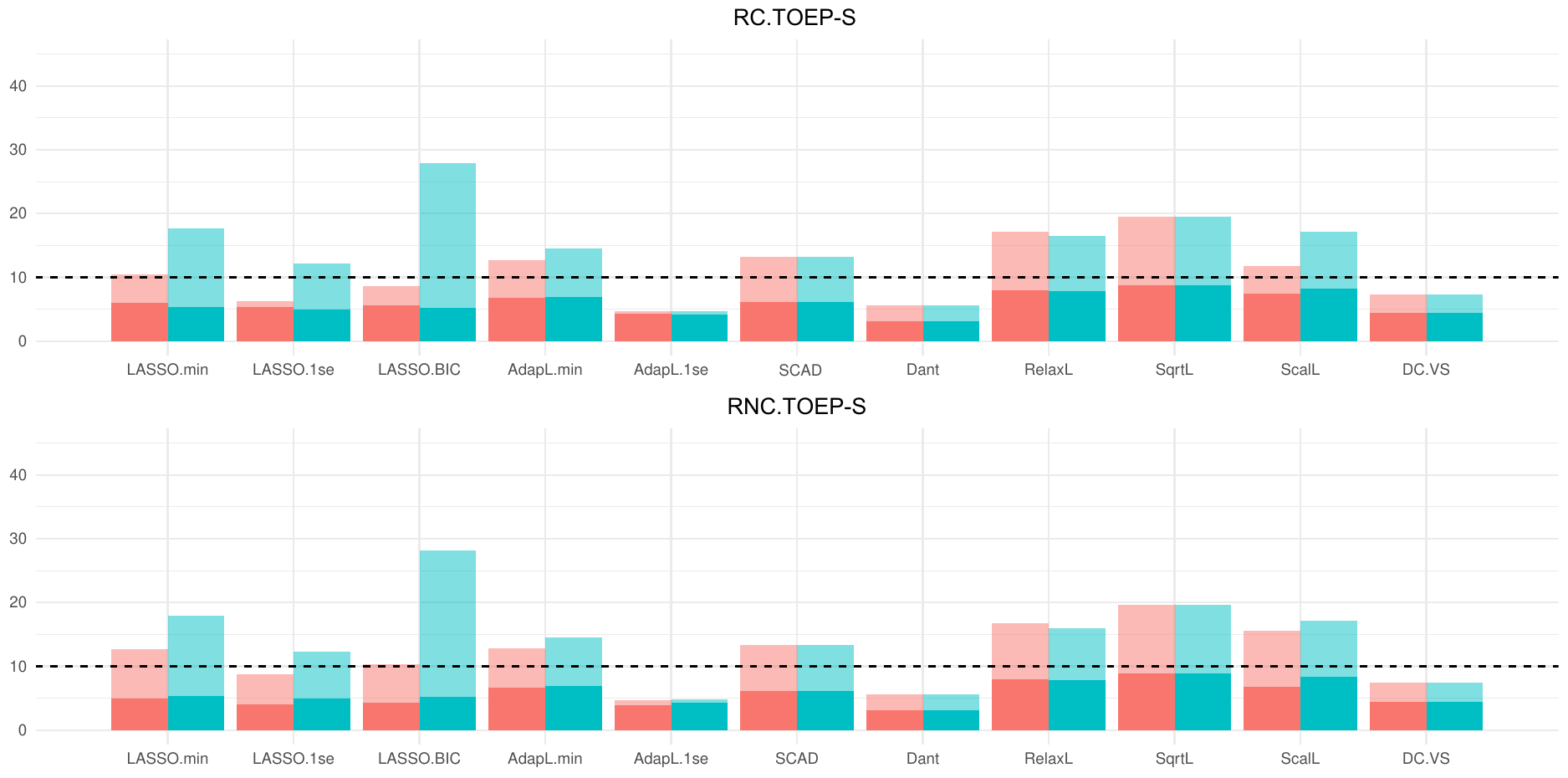}
	\caption{\label{cov_selected_comparison_LASSO_3a_3b_rho_0.9}Number of important covariates (dark \textcolor{coralred}{left}/\textcolor{my_seagreen}{right} rectangular area) and noisy ones (soft \textcolor{lightcoral}{left}/\textcolor{blue-green}{right} rectangular area) for proposed algorithms taking $p=100$ and selected in terms of $\mathcolor{coralred}{\mathbf{X}^{\text{r}}}$/$\mathcolor{my_seagreen}{\mathbf{X}^{\text{us}}}$ in scenarios RC.TOEP-S (the first row) and RNC.TOEP-S (the second row) for $\rho=0.9$ and $n=300$. The dashed line marks the $s=10$ value.}
\end{figure}

Finally, we move to the TOEP-S scenarios (RC.TOEP-S and RNC.TOEP-S). Results for $\rho=0.9$ are displayed in Figure \ref{cov_selected_comparison_LASSO_3a_3b_rho_0.9} and Table \ref{summ_scenario_3}. Those for $\rho=0.5$ are collected in Figure 37 and Table 24 in Section D.3 of the Appendix. As in the $p>n$ context (see Figure \ref{cov_selected_comparison_LASSO_3a_3b_rho_0.9_n_50}), all algorithms perform similarly in RC.TOEP-S as well as RNC.TOEP-S, except for including a bit more noise in this last scenario. Besides, both standardizations play a similar role for the AdapL.1se, SCAD, Dant, SqrtL and DC.VS performance. In the remaining algorithms, but for the RelaxL (LASSO.min, LASSO.1se, LASSO.BIC, AdapL.min and ScalL), the $\mathbf{X}^{\text{us}}$ approach selects more noisy covariates. Only the RelaxL decreases the noisy terms using $\mathbf{X}^{\text{us}}$. This is an example of how univariate standardization is not always the best option when there are dependence relations between covariates in different scales. Again, some procedures try a complete recovery of $S$ (LASSO.min, LASSO.1se, LASSO.BIC, AdapL.min, SCAD, RelaxL, SqrtL, ScalL and DC.VS for $\rho=0.5$), while others make use of the dependence structure selecting fewer covariates (AdapL.1se, Dant and DC.VS for $\rho=0.9$). While the first group obtains a better proportion of true positives, this adds quite noisy terms in exchange, especially in the $\rho=0.5$ case. Conversely, the AdapL.1se, Dant and DC.VS for $\rho=0.9$ choose fewer covariates but verify that the majority are important ones, although some of these are interchanged by unimportant terms pretty correlated in the $\rho=0.9$ case (see Figure 39). Into words, these are focused on reducing the false discovery rate. As it was already pointed out in the $p>n$ case, noticing that an efficient number of covariates needed to explain a great percentage of variability is less than $s=10$ in these scenarios (see Table 15 in Section B of the Appendix), this last strategy also makes sense. Additionally, these three procedures tend to pick up the important covariates with the greatest scales ($j=15$, $18$, $21$, $24$, $27$, $30$) a high percentage of times. We refer to Figures 38 and 39 in Section D.3 of the Appendix. In fact, the $n>p$ context helps to avoid the selection of unimportant covariates in higher scales than relevant ones when there are strong dependencies, as in the $\rho=0.9$ case, in comparison to the $p>n$ framework (see Figures 39 and 36 in Section D.3 of the Appendix, respectively).

\begin{table}[htb] 
	\centering
	\small 
	\adjustbox{max width=\textwidth}{
		\begin{tabular}{c cc cc cc cc}
			\toprule
			& \multicolumn{4}{c}{RC.TOEP-S} & \multicolumn{4}{c}{RNC.TOEP-S} \\
			\cmidrule(rl){2-5} \cmidrule(rl){6-9}
			& \multicolumn{2}{c}{\textbf{RAW}} &  \multicolumn{2}{c}{\textbf{UNIV.}} & \multicolumn{2}{c}{\textbf{RAW}} &  \multicolumn{2}{c}{\textbf{UNIV.}} \\
			\cmidrule(r){2-3}  \cmidrule(rl){4-5} \cmidrule(rl){6-7} \cmidrule(rl){8-9} 
			\vspace{-0.07cm}
			\rule{0pt}{0.35cm} \multirow{2}{*}{ \textbf{METHOD}} & {\footnotesize \textbf{MSE}} &  {\footnotesize \textbf{\% Dev}} & {\footnotesize \textbf{MSE}} &  {\footnotesize \textbf{\% Dev}} & {\footnotesize \textbf{MSE}} &  {\footnotesize \textbf{\% Dev}} & {\footnotesize \textbf{MSE}} &  {\footnotesize \textbf{\% Dev}} \\
			& {\scriptsize$(7.818)$} & {\scriptsize$(0.9)$} & {\scriptsize$(7.818)$} & {\scriptsize$(0.9)$} & {\scriptsize$(7.818)$} & {\scriptsize$(0.9)$} & {\scriptsize$(7.818)$} & {\scriptsize$(0.9)$} \\ 
			\cmidrule(r){1-3} \cmidrule(rl){4-5} \cmidrule(rl){6-7} \cmidrule(rl){8-9} 
			\rule{0pt}{0.35cm} \textbf{LASSO.min} & \cellcolor[gray]{.9}7.100 & 0.908 & 6.920 & 0.910 & 7.022 & 0.909 & 6.926 & 0.911 \rule[-0.25cm]{0pt}{0pt} \\
			\textbf{LASSO.1se} & \cellcolor[gray]{.9}7.561 & 0.902 & \cellcolor[gray]{.9}7.302 & 0.906 & \cellcolor[gray]{.9}7.563 & 0.902 & \cellcolor[gray]{.9}7.302 & 0.906 \rule[-0.25cm]{0pt}{0pt} \\
			\textbf{LASSO.BIC} & \cellcolor[gray]{.9}7.643 & 0.901 & \cellcolor[gray]{.9}7.527 & 0.903 & \cellcolor[gray]{.9}7.546 & 0.903 & \cellcolor[gray]{.9}7.526 & 0.903 \rule[-0.25cm]{0pt}{0pt} \\ 
			\textbf{AdapL.min} & \cellcolor[gray]{.9}7.060 & 0.909 & 6.920 & 0.91 & \cellcolor[gray]{.9}7.052 & 0.909 & 6.919 & 0.911 \rule[-0.25cm]{0pt}{0pt} \\   
			\textbf{AdapL.1se} & \squaretext{8.924} & 0.885 & \squaretext{8.790} & 0.886 & \squaretext{8.936} & 0.885 & \squaretext{8.773} & 0.887 \rule[-0.25cm]{0pt}{0pt} \\ 
			\textbf{SCAD} & \cellcolor[gray]{.9}7.149 & 0.907 & \cellcolor[gray]{.9}7.149 & 0.907 & \cellcolor[gray]{.9}7.146 & 0.908 & \cellcolor[gray]{.9}7.146 & 0.908 \rule[-0.25cm]{0pt}{0pt} \\ 
			\textbf{Dant} & \squaretext{11.804} & 0.847 & \squaretext{11.804} & 0.847 & \squaretext{11.906} & 0.846 & \squaretext{11.906} & 0.846 \rule[-0.25cm]{0pt}{0pt} \\ 
			\textbf{RelaxL} & \cellcolor[gray]{.9}7.182 & 0.907 & \cellcolor[gray]{.9}7.221 & 0.907 & \cellcolor[gray]{.9}7.197 & 0.907 & \cellcolor[gray]{.9}7.245 & 0.906 \rule[-0.25cm]{0pt}{0pt} \\  
			\textbf{SqrtL} & \cellcolor[gray]{.9}7.192 & 0.907 & \cellcolor[gray]{.9}7.192 & 0.907 & \cellcolor[gray]{.9}7.186 & 0.907 & \cellcolor[gray]{.9}7.186 & 0.907 \rule[-0.25cm]{0pt}{0pt} \\  
			\textbf{ScalL} & \cellcolor[gray]{.9}7.337 & 0.905 & \cellcolor[gray]{.9}7.184 & 0.907 & \cellcolor[gray]{.9}7.259 & 0.906 & \cellcolor[gray]{.9}7.181 & 0.907  \rule[-0.25cm]{0pt}{0pt} \\ 
			\textbf{DC.VS} & \cellcolor[gray]{.9}\squaretext{8.059} & 0.896 & \cellcolor[gray]{.9}\squaretext{8.059} & 0.896 & \cellcolor[gray]{.9}\squaretext{8.050} & 0.896 & \cellcolor[gray]{.9}\squaretext{8.050} & 0.896 \rule[-0.25cm]{0pt}{0pt} \\  
			\bottomrule
		\end{tabular}
	}
	\caption[]{Comparison of all proposed algorithms for $p=100$, $n=300$ and $\rho=0.9$ using $\mathbf{X}^{\text{r}}$ and $\mathbf{X}^{\text{us}}$ in Scenario 3. Oracle values are in brackets. \hlc[gray!20]{Highlighted} terms are values in  \small{$[0.9\cdot\text{MSE},1.1\cdot\text{MSE}]$} (GI) and \dashbox[1.2cm][c]{squared} ones those that correct overestimation. }
	\label{summ_scenario_3}
\end{table}

Lastly, prediction results for RC.TOEP-S and RNC.TOEP-S are displayed in Table \ref{summ_scenario_3} for $\rho=0.9$ and in Table 24 in Section D.3 of the Appendix. Only the AdapL.1se and the Dant correct the overestimation in all cases. We can also add the DC.VS to this list in the $\rho=0.9$ case. For $\rho=0.5$, we can see that LASSO.1se with $\mathbf{X}^{\text{r}}$, LASSO.BIC, AdapL.1se with $\mathbf{X}^{\text{us}}$, RelaxL, SqrtL, ScalL with $\mathbf{X}^{\text{r}}$ and DC.VS obtain a MSE value in the GI for RC.TOEP-S. This is also verify in the RNC.TOEP-S scenario for the same procedures, except for ScalL with $\mathbf{X}^{\text{r}}$. In terms of $\rho=0.9$, this condition is satisfied in both scenarios for LASSO.1se, LASSO.BIC, AdapL.min with $\mathbf{X}^{\text{r}}$, SCAD, RelaxL, SqrtL, ScalL and DC.VS. The LASSO.min with $\mathbf{X}^{\text{r}}$ is also added to this list for the RC.TOEP-S scenario with $\rho=0.9$.

\subsection{Comparison with thresholding techniques}\label{Comparison_thr}

As mentioned in Section \ref{intro}, apart from covariates selection techniques relying on some underlying structure or using dependence coefficients, one could resort to thresholding techniques. At this point, one can wonder how these procedures would perform for different dependence-scales structures. For this aim, we test the performance of these algorithms. As a result, we try to sort covariates' relevance to establish a proper cutoff or threshold to define a first screening step. For this aim, the performance of the coefficient of determination R$^2$ (see, for example, \cite{Glantz1990}), the distance correlation coefficient of \cite{Szekely2007} (DC), and partial least squares of \cite{Wold1966} (PLS) values are used as measures of relevance for scenarios RNC.IND, UTOEP-S and RNC.TOEP-S introduced in Section \ref{simulation_scenarios}. Complete results and an extended analysis are collected in Section C of the Appendix.   

\begin{figure}[htb]\centering
	\includegraphics[width=\linewidth]{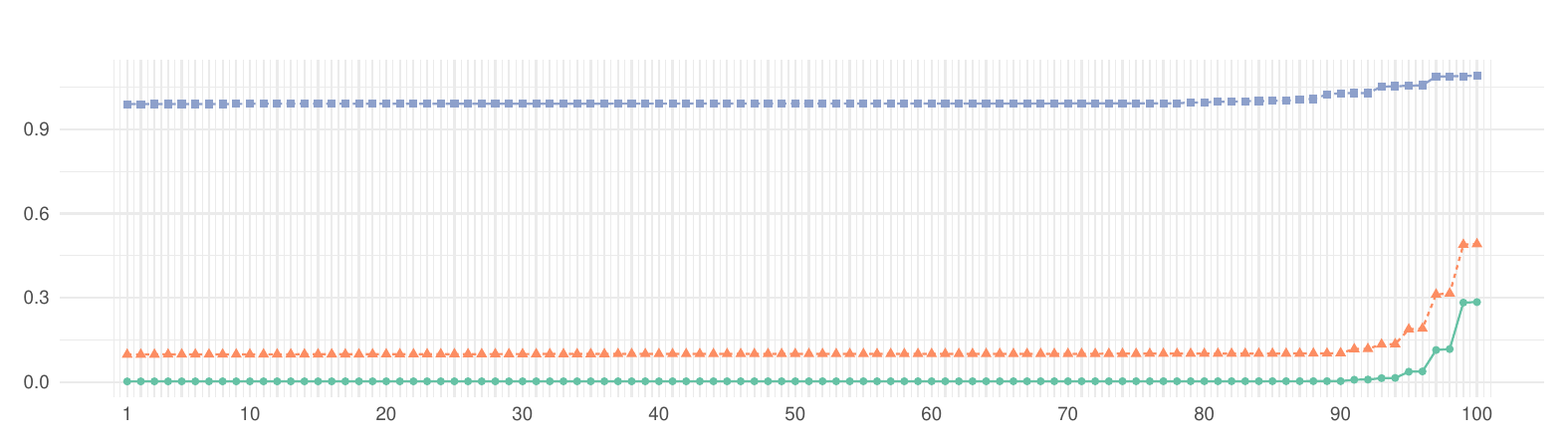}
	\caption{\label{plot_ordered_mean_values_1c}Representative covariates relevance coefficients in increasing order in terms of R$^2$ (\textcolor{junglegreen}{\large $\bullet$}), DC (\textcolor{lightcoral}{$\blacktriangle$}) and PLS (\textcolor{bluegray}{\scriptsize $\blacksquare$}) coefficients in RNC.IND scenario using $\mathbf{X}^{\text{r}}$.}
\end{figure}

There, we can see as the R$^2$, as well as the DC coefficients, are good options for a first screening step which helps to clean the data. Nevertheless, screening techniques based on these ideas should be followed for a second step, using some covariates selection algorithm. Otherwise, so much noise would be added to the model trying to recover $S$. Even under independence, as in the RNC.IND scenario, one can see as it is difficult to establish an optimal cutoff without noise addition (see Figure \ref{plot_ordered_mean_values_1c}). This phenomenon is intensified when there exist dependence structures (see Figures 16 and 17 in Section C of the Appendix). As a result, screening procedures also suffer from dependence and scale effects no matter the employed threshold. Hence, these have similar limitations to penalization techniques or dependence coefficients under dependence structures and/or covariates in different scales. Additionally, it is worth noting that the R$^2$ coefficient is only applicable to linear structures and has an advantage under the normality assumption. If we guess other types of relations, other approaches like the DC coefficient are more suitable. Related to the PLS approach, this has displayed a really bad performance in practice, being unable to discriminate between relevant and noisy covariates even in the RNC.IND framework (see Figure \ref{plot_ordered_mean_values_1c}). Thus, we do not recommend its use in practice.

\section{Real datasets}\label{data}

In this section, the performance of the different procedures considered along Section \ref{competitors} are tested in three real datasets: riboflavin, body fat and Portuguese wine. These data are examples of problems where different dependence structures and scale effects arise. Next, these are briefly introduced and analyzed. We refer the reader to Section E of the Appendix for more details\footref{foot2}.

\subsection{Riboflavin}\label{sec:riboflavin_main}

Riboflavin\footnote{The \texttt{riboflavin} dataset is available in library \texttt{hdi} (\cite{hdi}) of \cite{R}} is a high-dimensional genomic dataset where $p>n$. This contains a total of $n=71$ samples. Each sample measures the logarithm of the expression level of $p=4088$ genes that are believed to be related to the rate of riboflavin (vitamin B2) production of the \textit{Bacillus subtilis} bacterium. This dataset has already been studied in works as the one of \cite{Buhlmann2014}. In this example, all covariates seem to have a similar range of scale values (see Figure 42 in Section E.1 of the Appendix), but there are different types of dependence patterns between them (see Figure 43 in Section E.1 of the Appendix).

In practice, this data has been centered to avoid the intercept in the model without loss of generality. Although it seems reasonable to assume that all covariates are on a similar scale, small differences can been appreciated in Figure 42 in Section E.1 of the Appendix. As a result, we have considered $\mathbf{X}^{\text{r}}$ as well as $\mathbf{X}^{\text{us}}$ approaches.

\begin{figure}[htb]\centering
	\includegraphics[width=\linewidth]{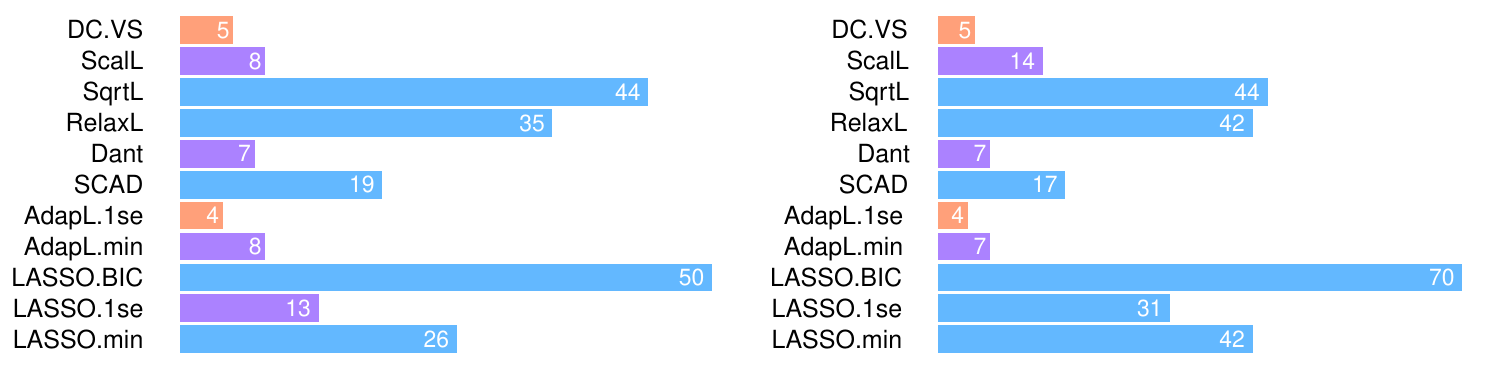}
	\caption{\label{barplots_riboflavin_ggplot}Number of selected covariates for the considered procedures for $\mathbf{X}^{\text{r}}$ (left) and $\mathbf{X}^{\text{us}}$ (right) of the riboflavin dataset.}
\end{figure}

The number of selected covariates for each of the eleven considered algorithms is displayed in Figure \ref{barplots_riboflavin_ggplot}. There are pretty differences between the number of selected terms for these procedures. The LASSO.BIC is the one that selects the greatest number of terms, as expected for a $p>n$ framework (see results in Sections \ref{LASSO_results} and \ref{competitors}). This applies in both, $\mathbf{X}^{\text{r}}$ and $\mathbf{X}^{\text{us}}$ versions. The SqrtL, RelaxL, LASSO.min, SCAD (under $\mathbf{X}^{\text{r}}$) and LASSO.1se follow this. As mentioned in Sections \ref{LASSO_results} and \ref{competitors}, it is expected for these algorithms in this type of contexts to recover a larger part of the unknown $S$ set. In exchange, these would add quite a noise in the process. In contrast, AdapL.1se, DC.VS and Dant are the algorithms that select fewer covariates. These last are more conservative in this context, guaranteeing that a high percentage of the selected covariates are relevant. Nevertheless, some important terms could be exchanged with irrelevant ones when having $p>n$, strong dependency structures and covariates with different scales. We refer to Section \ref{competitors} analysis for more insight.

The genes selected by the $\mathbf{X}^{\text{r}}$ and $\mathbf{X}^{\text{us}}$ approaches are displayed in Tables 28 and 29 in Section E.4.1 of the Appendix, respectively. Most of the procedures change their selection in terms of the standardization version employed. Only the Dant, SqrtL and the DC.VS algorithms keep their selection. Besides, the Dant, AdapL.1se and the AdapL.min (under $\mathbf{X}^{\text{us}}$) select a large percentage of popular genes, which means those most selected by the eleven studied algorithms.

\cite{Buhlmann2014} apply stability selection with randomized LASSO over the riboflavin data and detect three stable genes: \texttt{LYSC\_at}, \texttt{YOAB\_at}, and \texttt{YXLD\_at}. The most similar selection is the one performed by the adaptive versions AdapL.min and AdapL.1se, as well as the Dant selector, using $\mathbf{X}^{\text{us}}$. These techniques select 7, 4 and 7 genes, respectively, out of the $p=4088$ available. Their selections include \texttt{YOAB\_at} and \texttt{YXLD\_at} genes. The \texttt{LYSC\_at} gen is selected by the RelaxL and SqrtL, as well as the LASSO versions (LASSO.min, LASSO.1se and LASSO.BIC) and the ScalL using $\mathbf{X}^{\text{us}}$. As it has been seen for different examples in Sections \ref{LASSO_results} and \ref{competitors}, this selection could be due to spurious correlations: all the mentioned algorithms tend to improve the true positive proportion paying the price of noise addition. However, this gen may be a true relevant one, but a greater sample size could be needed for its recovery. Furthermore, other important genes could be avoided in the \cite{Buhlmann2014} selection because of the same reason. Some possible candidates would be those repeated a great number of times or the ones selected for procedures that have displayed a more robust behavior in this type of scenarios in the simulations, such as the AdapL.1se, Dant or the DC.VS. 

In conclusion, although a larger sample size would be necessary to ensure an adequate recovery of the relevant terms, the covariates selection algorithms for the context $p>n$ allow a great dimensionality reduction. This transforms the problem into a tractable one. From $p=4088$ genes, all algorithms select fewer than $n=71$ terms. Besides, one can work with fewer than 10 covariates using the selections of the AdapL.1se, Dant or DC.VS approaches, among others. These procedures correspond to the ones that tend to select a representative subset of the $s$ relevant terms using the different scales and dependence structure of the data. These are also the ones that get the smallest percentage of noisy terms. Instead, if one wants to get the maximum possible recovery of relevant genes regardless of the inclusion of noise, algorithms such as LASSO.1se, RelaxL, SqrtL and ScalL would be more suitable.

\begin{table}[htb] 
	\centering
	\adjustbox{max width=\textwidth}{
	\small 
	\begin{tabular}{c cccccc}
		\toprule  
		& LASSO.min & LASSO.1se & LASSO.BIC & AdapL.min & AdapL.1se & SCAD\\ 
		\cmidrule{2-7}
		\rule{0pt}{0.4cm} \textbf{MSE} & 0.121 & 0.128 & 0.517 & 0.092 & 0.161 & 0.082 \\
		\rule{0pt}{0.4cm} \textbf{\% Dev} & 0.849 & 0.838 & 0.372 & 0.883 & 0.798 & 0.897 \rule[-0.35cm]{0pt}{0pt}\\ 
		\hline
		\rule{0pt}{0.4cm} & Dant & RelaxL & SqrtL & ScalL & DC.VS \\
		\cmidrule{2-6}
		\rule{0pt}{0.4cm} \textbf{MSE} & 0.223 & 0.116 & 0.670 & 0.194 & 0.234 \\
		\rule{0pt}{0.4cm} \textbf{\% Dev} & 0.741 & 0.852 & 0.141 & 0.767 & 0.732 \\
		\bottomrule
	\end{tabular}
	}
	\caption{Estimation results of the riboflavin dataset using a training sample of size 55 and a testing one of 16 taking $\mathbf{X}^{\text{us}}$ in 100 simulations.}
	\label{estimation_riboflavin_univ}
\end{table}

Concerning predictions, we apply a four-step procedure. First, the relevant covariates are selected employing the total $n=71$ samples. Subsequently, the sample is randomly divided in two: a training sample of size 55 (approx. 80\%) and a testing one of 16 (approx. 20\%)\footnote{Notice that, in this case, the LASSO.BIC using $\mathbf{X}^{\text{us}}$ selects 70 covariates when employing the $n=71$ samples. As 70 is greater than the 55 elements of the training sample, it is not possible to directly adjust the linear regression model because of the $p>n$ situation. To address this issue, we apply a new relevant covariates selection step using the training data for each simulation and selecting a maximum of 55 elements.}. Finally, we adjust a linear regression model using the training sample and just the covariates previously selected. Next, we test its prediction performance using the testing sample. For this purpose, we compute the MSE and the \% Dev (see Section \ref{simulation_scenarios} for more insight). This process is repeated in a total of 100 simulations. Results for $\mathbf{X}^{\text{us}}$ are displayed in Table \ref{estimation_riboflavin_univ}, while those concerning $\mathbf{X}^{\text{r}}$ are collected in Table \ref{estimation_riboflavin_univ} in Section E.4.1 of the Appendix.

We can appreciate in Table \ref{estimation_riboflavin_univ}, and Table 27 in Section E.4.1 of the Appendix, as MSE related to the procedures which have displayed a good performance correcting the overestimation under dependence and covariates in different scales (AdapL.1se, Dant and DC.VS) tend to be larger than the other ones. We can also add to this group the LASSO.BIC, RelaxL and SqrtL because their MSE values tend to be in the GI in this type of scenarios when the number of relevant covariates selected to adjust the model is less than $n$ (see Figure \ref{barplots_riboflavin_ggplot}). One can guess that the remaining procedures overestimate the prediction results.

\subsection{Body fat}\label{sec:bodyfat}

The body fat dataset\footnote{See \url{https://www.kaggle.com/datasets/fedesoriano/body-fat-prediction-dataset}} consists of 14 body measures taken in 252 men. The aim is to determine the percentage of body fat using these covariates: underwater weighing or density, age (years), weight (lbs), height (inches), and neck, chest, abdomen, hip, thigh, knee, ankle, biceps (extended), forearm as well as wrist circumference (cm). The body fat variable is obtained using Siri's equation $\text{BodyFat}=4.95/\text{Density} - 4.50$ (\cite{Siri1956}).

\begin{figure}[htb]\centering
	\includegraphics[width=\linewidth]{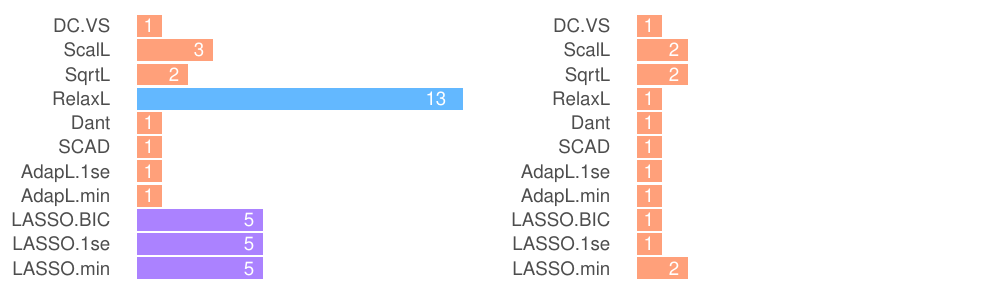}
	\caption{\label{barplots_bodyfat_ggplot}Number of selected covariates for the considered procedures for $\mathbf{X}^{\text{r}}$ (left) and $\mathbf{X}^{\text{us}}$ (right) of the body fat dataset.}
\end{figure}

First of all, we apply Box-Cox transformations in all $X_j$+0.01 and $Y$+0.01 variables to avoid skewness\footnote{\label{foot:box-cox}The \texttt{boxcox} function of the \texttt{MASS} library of R (\cite{R}) is used.}, for $j=1,\dots,14$. Next, we remove possible outliers using the Mahalanobis distance\footnote{\label{foot:outliers}We employ the \texttt{mdepth.MhD} function of the library \texttt{fda.usc} (\cite{fda.usc}).}. The $5\%$ of less depth samples are removed. The resulting variables are displayed in Figure 44 in Section E.2 of the Appendix. This processed data has a sample size of length $n=239$ and $p=14$ covariates, translating in a $n\geq p$ framework\footnote{\label{foot:GitHub_data}The processed  dataset can be found in the GitHub repository \url{https://github.com/LauraFreiG/Covariates_selection.git}. In particular, in the ``Real data sets'' folder, inside the ``Linear Regression'' one.}. This dataset has covariates highly correlated between them as it is displayed through its correlation matrix values (see Figure 45 in Section E.2 of the Appendix). Additionally, these are in quite different scales (see Table 25 in Section E.2 of the Appendix). As a result, this is an example of strong dependence structures and covariates in different scales in a $n\geq p$ framework. Again, we work with the centered variables and analyze the results for both $\mathbf{X}^{\text{r}}$ and $\mathbf{X}^{\text{us}}$.

The number of covariates selected by each of the studied algorithms is showed in Figure \ref{barplots_bodyfat_ggplot}. In this example, all procedures tend to pick fewer or the same number of covariates applying $\mathbf{X}^{\text{us}}$. This fact may be motivated by the notable differences in the scale values of the covariates. AdapL.min, AdapL.1se, SCAD, Dant and DC.VS algorithms are the only procedures that keep selecting just one covariate in both frameworks. In the $\mathbf{X}^{\text{us}}$ case, all algorithms select less or equal to 2 covariates. This may be because of the existence of strong correlations between covariates. Thus, just a bunch of them may explain all the information. In contrast, for the $\mathbf{X}^{\text{r}}$ case, only the AdapL.min, AdapL.1se, SCAD, Dant and SqrtL select fewer than 3 terms. Given the results of Section \ref{LASSO_results}, we see that the AdapL.1se, the Dant and the DC.VS tend to be more conservative in this type of scenarios. In other words, these select fewer covariates but guarantee that these are relevant with a high probability. This behavior contrasts with procedures such as the RelaxL, ScalL or SqrtL. These select more features for a proper complete recovery of $S$ but add quite a noise in exchange.

Covariates selected for each algorithm are collected in Table 31 in Section E.4.2 of the Appendix. The terms most selected for both types of standardizations are density and age. We have to also add forearm in the $\mathbf{X}^{\text{r}}$ case. In particular, in the $\mathbf{X}^{\text{us}}$ approach, several procedures (LASSO.1se, LASSO.BIC, AdapL.min, AdapL.1se, SCAD, Dant and RelaxL) select only the density covariate. Therefore, noticing the Siri's equation, a possible interpretation is that body fat is approximately well-explained using a linear expression considering only the density term.

This is a real example where the use of standardization methods—or the lack thereof—can produce significant differences. While $\mathbf{X}^{\text{us}}$ helps to mitigate the scale effects, it can also distort the correlation structure, nulling its effects. This distortion might account for the differences observed in the selections made by some of the algorithms considering $\mathbf{X}^{\text{r}}$ and $\mathbf{X}^{\text{us}}$ approaches. See Section \ref{competitors} for more insights. Only five procedures keep the same selection for both standardization versions: the AdapL.min, SCAD, Dant, SqrtL and DC.VS. The Dant, jointly with the AdapL.1se, have displayed the best results in terms of guaranteeing that almost all selected covariates are relevant. In contrast, if one wants to verify that a larger part of $S$ is recovered, other procedures are more suitable for this purpose. In this last case, the addition of noise tends to be inevitable. We refer the reader to Section \ref{competitors}.

\begin{table}[htb] 
	\centering
	\adjustbox{max width=\textwidth}{
		\small 
		\begin{tabular}{c cccccc}
			\toprule  
			& LASSO.min & LASSO.1se & LASSO.BIC & AdapL.min & AdapL.1se & SCAD\\ 
			\cmidrule{2-7}
			\rule{0pt}{0.4cm} \textbf{MSE} & 0.059 & 0.059 & 0.059 & 0.059 & 0.059 & 0.059  \\
			\rule{0pt}{0.4cm} \textbf{\% Dev} & 0.997 & 0.997 & 0.997 & 0.997 & 0.997 & 0.997 \rule[-0.35cm]{0pt}{0pt}\\ 
			\hline
			\rule{0pt}{0.4cm} & Dant & RelaxL & SqrtL & ScalL & DC.VS \\
			\cmidrule{2-6}
			\rule{0pt}{0.4cm} \textbf{MSE} & 0.059 & 0.059 & 0.059 & 0.059 & 0.059 \\
			\rule{0pt}{0.4cm} \textbf{\% Dev} & 0.997 & 0.997 & 0.997 & 0.997 & 0.997 \\
			\bottomrule
		\end{tabular}
	}
	\caption{Estimation results of the body fat dataset using a training sample of size 191 and a testing one of 48 taking $\mathbf{X}^{\text{us}}$ in 100 simulations.}
	\label{estimation_bodyfat_univ}
\end{table}

Related to predictions, we apply a similar procedure as the one introduced above in Section \ref{sec:riboflavin_main}. First, we detect the relevant covariates for each algorithm. Then, we randomly consider a training sample about the 80\% of data (191 samples) and a testing one of the 20\% (48 terms) and compute their associated MSE and \% Dev. We repeat this last procedure in 100 repetitions. Results for the $\mathbf{X}^{\text{r}}$ case are displayed in Table 30 in Section E.4.2 of the Appendix. Those for $\mathbf{X}^{\text{us}}$ are collected in Table \ref{estimation_bodyfat_univ}. In this type of scenarios, almost all procedures tend to correct the overestimation or verify that their MSE is in the GI (see Section \ref{competitors}). Results are pretty similar using the $\mathbf{X}^{\text{r}}$ approach, except for the ones of AdapL.1se and ScalL. This last can be motivated by a bad selection of the relevant terms. An explanation for the adaptive versions is the scales effects using $\mathbf{X}^{\text{r}}$, as these algorithm completely change their selection using $\mathbf{X}^{\text{us}}$. Taking $\mathbf{X}^{\text{us}}$, all procedures tend to select just density (with 1 or 3 extra ones in some cases), so all prediction results are very similar.

\subsection{Portuguese wine}

The Portuguese red wine dataset\footnote{This is available in http://www3.dsi.uminho.pt/pcortez/wine/} contains several physicochemical
parameters about 1599 samples of the red vinho verde, which is a typical wine type from the northwest regions of Portugal. These parameters are fixed acidity, volatile acidity, citric acid, residual sugar, chlorides, free sulfur dioxide, total sulfur dioxide, density, pH, sulphates and alcohol. See \cite{Cortez2009} for more details. The objective is to model the alcohol by volume content using the rest of the 10 variables.

Similar to the body fat dataset, we apply Box-Cox transformations\footref{foot:box-cox} of the variables and clean possible outliers\footref{foot:outliers}. We refer to Section \ref{sec:bodyfat} or Section E.3 of the Appendix for more details. This results in a total of $n=1519$ samples and $p=10$ covariates\footref{foot:GitHub_data} (see Figure 46 in Section E.3 of the Appendix). This is a $n\geq p$ example where there exists some, but not too much, strong dependence relations (see Figure 47 in Section E.3 of the Appendix). Besides, all covariates have a scale in a similar range of values (see Table 26 in Section E.3 of the Appendix). One more time, we centered the data without loss of generality and consider $\mathbf{X}^{\text{r}}$ and $\mathbf{X}^{\text{us}}$ versions.

\begin{figure}[htb]\centering
	\includegraphics[width=\linewidth]{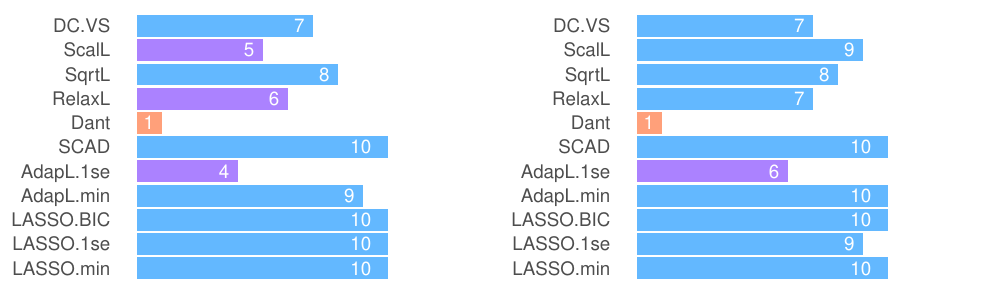}
	\caption{\label{barplots_red_wine_ggplot}Number of selected covariates for the considered procedures for $\mathbf{X}^{\text{r}}$ (left) and $\mathbf{X}^{\text{us}}$ (right) of the Portuguese wine dataset.}
\end{figure}

The number of covariates selected by the studied algorithms is displayed in Figure \ref{barplots_red_wine_ggplot}. As expected in this case, results do not differ too much between $\mathbf{X}^{\text{r}}$ and $\mathbf{X}^{\text{us}}$ approaches. All algorithms keep the same number of relevant terms or this tends to vary in one or two units, but for the ScalL. LASSO.min, LASSO.BIC, SCAD, Dant and SqrtL and DC.VS keep the same quantity in both, while this decreases for the remaining algorithms in the $\mathbf{X}^{\text{us}}$ case (except for AdapL.min, AdapL.1se, RelaxL and ScalL). All procedures add at least seven covariates, but for AdapL.1se, Dant, and RelaxL as well as ScalL in the $\mathbf{X}^{\text{r}}$ case.  

The covariates selected by each procedure are collected in Table 33 in Section E.4.3 of the Appendix. Although the number of selected terms is similar for $\mathbf{X}^{\text{r}}$ and $\mathbf{X}^{\text{us}}$ versions, only the LASSO.min, LASSO.BIC, SCAD, Dant, RelaxL, SqrtL and DC.VS keep the same selection of covariates in both. Total sulfur, sulphates, density, sugar, citric acidity and fixed acidity are the covariates selected by most algorithms for both types of standardized versions. These are followed by chlorides and pH. The covariates most times excluded are the volatile acidity and free sulfur. Then, one can see that, in global terms, selection results are quite similar between both standardizations. Nevertheless, the covariates' relevance, in terms of the percentage of times selected, changes from one procedure to another. AdapL.1se using $\mathbf{X}^{\text{us}}$ and DC.VS are the procedures which select most of these popular terms. Furthermore, AdapL.1se, Dant and DC.VS, have displayed to be the most optimal ones for minimizing the false discovery proportion in this type of scenarios (see Section \ref{competitors} fro more details). As a result, one can trust in the AdapL.1se using $\mathbf{X}^{\text{us}}$, Dant or DC.VS selection to have more guarantees of only recovering relevant terms. On the contrary, procedures such as RelaxL, SqrtL or ScalL are more prone to recover as many important terms as possible in this framework (we refer to Section \ref{competitors}). Some noisy covariates tend to be selected as a trade-off in this last case.

\begin{table}[htb] 
	\centering
	\adjustbox{max width=\textwidth}{
		\small 
		\begin{tabular}{c cccccc}
			\toprule  
			& LASSO.min & LASSO.1se & LASSO.BIC & AdapL.min & AdapL.1se & SCAD\\ 
			\cmidrule{2-7}
			\rule{0pt}{0.4cm} \textbf{MSE} & $<10^{-6}$ & $<10^{-6}$ & $<10^{-6}$ & $<10^{-6}$ & $<10^{-6}$ & $<10^{-6}$ \\
			\rule{0pt}{0.4cm} \textbf{\% Dev} & 0.668 & 0.666 & 0.668 & 0.668 & 0.643 & 0.668 \rule[-0.35cm]{0pt}{0pt}\\ 
			\hline
			\rule{0pt}{0.4cm} & Dant & RelaxL & SqrtL & ScalL & DC.VS \\
			\cmidrule{2-6}
			\rule{0pt}{0.4cm} \textbf{MSE} & $<10^{-6}$ & $<10^{-6}$ & $<10^{-6}$ & $<10^{-6}$ & $<10^{-6}$ & \\
			\rule{0pt}{0.4cm} \textbf{\% Dev} & 0.238 & 0.591 & 0.666 & 0.667 & 0.665 & \\
			\bottomrule
		\end{tabular}
	}
	\caption{Estimation results of the Portuguese wine dataset using a training sample of size 191 and a testing one of 48 taking $\mathbf{X}^{\text{us}}$ in 100 simulations.}
	\label{estimation_wine_univ}
\end{table}

For prediction results, we replicate the procedure introduced above in Sections \ref{sec:riboflavin_main} and \ref{sec:bodyfat}. In this case, the training sample is about 1215 samples and the testing one of 304. Obtained MSE and \% Dev for $\mathbf{X}^{\text{r}}$ are collected in Table 32 in Section E.4.3 of the Appendix and those for $\mathbf{X}^{\text{us}}$ are displayed in Table \ref{estimation_wine_univ}.

 In terms of $\mathbf{X}^{\text{r}}$, the LASSO.min, LASSO.1se, LASSO.BIC, AdapL.min, AdapL.1se, Dant and ScaL obtain an MSE greater than the other procedures. This changes using $\mathbf{X}^{\text{us}}$. As expected in view of the covariates selection, results are similar for both $\mathbf{X}^{\text{r}}$ and $\mathbf{X}^{\text{us}}$ cases. Only the Dant obtains a significative lower \% Dev in comparison with the rest of the values for  $\mathbf{X}^{\text{us}}$ (this is the procedure which selects fewer covariates: just one). In contrast, we can also highlight the AdapL.min, AdapL.1se and the ScalL in the $\mathbf{X}^{\text{r}}$ framework. Algorithms like the AdapL.1se or Dant have displayed good properties in terms of avoiding overestimation in this type of scenarios, so we can use their associated MSE values as a guide.


\section{Discussion and guidelines}\label{discussion}

In a multivariate regression setup with $p>1$ covariates, a preliminary covariates selection step to consider only the relevant terms out of the $p$ candidates is desirable. This is particularly remarkable in high-dimensional regimes where $p>n$. Penalization techniques are an appealing alternative for this purpose, especially in $p>n$ frameworks (we refer to Section \ref{intro} for more details). One of the most well-known and still widely employed options is the LASSO procedure of \cite{Tibshirani1996}. Nevertheless, even in an easy framework as the orthogonal design, this procedure suffers from important limitations (see Sections \ref{intro}, \ref{scales_LASSO} and \ref{LASSO_results} for more details). One of these drawbacks concerns the covariates' scale effect (we refer the reader to Section \ref{scales_LASSO}). In practice, it is quite common to implement a univariate standardization step to prevent from this phenomenon. Nontheless, possible dependence structures can be disrupted, leading to wrong selections. As a result, it is interesting to know what to expect when working with non independent designs and covariates in different scales. Related to the dependence concern, previous studies in the literature have been carried out to arise some light. To say a few, one can see \cite{Meinshausen2010}, \cite{buhlmann2011statistics} or \cite{FreijeiroGonzalez2022}. However, to the best of our knowledge, no guidelines is provided in the literature when one has to face dependence relations jointly with covariates in different scales. To fill this gap we provide an extensive simulation studio considering scenarios with different dependence relations and configurations of the covariates scales. These scenarios are introduced in Section \ref{simulation_scenarios}. In Section \ref{LASSO_results}, we test the LASSO performance on these scenarios differentiating between working with the raw data (without standardization) or applying a first univariate standardization step. Next, we compare their performance with that of some competitors and analyze their performance in Section \ref{competitors}. Eventually, all these procedure are tested in some real datasets with different dependence-scales patterns in Section \ref{data}.

\begin{figure}[h!]\centering
	\hspace*{-1.1cm}
	\includegraphics[width=1.13\linewidth]{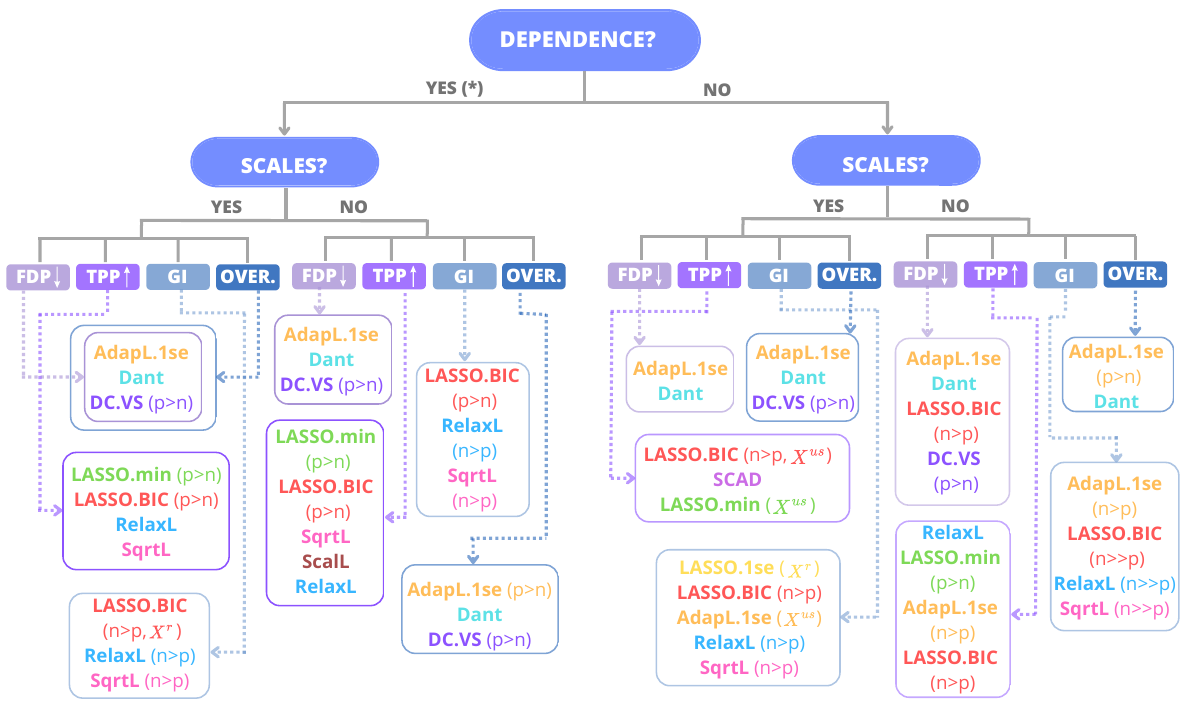}
	\caption{\label{scheme_options}Most adequate procedures to face dependence and/or scale effects to minimize the FDP (FDP$\downarrow$), to maximize the TPP (TPP$\uparrow$), to obtain an MSE in the Good Interval (GI) or to correct the overestimation (OVER.). Clarification between brackets means that the procedures only apply in those cases. (*) It depends on the type of dependence structure.}
\end{figure}

Summing up all the information provided for the simulation study, one can conclude that choosing a proper selector algorithm will depend on the practitioner objective as well as the working framework. First of all, one have to choose between searching for a proper recovery of the relevant terms, those in $S$, or searching for the covariates which improve the prediction capability of the model. These objectives are not compatible (see \cite{Yang2005} or \cite{Leng2006} among others). In terms of a suitable recovery of $S$, there are two possibilities: procedures which minimize the false discovery proportion (FDP) and those that search for a maximization of the true positive proportion (TPP). In the first group, the algorithms try to guarantee that all selected terms are relevant, although some terms of $S$ are missed. Conversely, algorithms in the second group search for the complete recovery of the $s$ relevant terms. In this last case, noise addition is allowed as a trade-off. Concerning prediction, the majority of procedures tend to overestimate the prediction capability. We refer to Sections \ref{LASSO_results} and \ref{competitors} for more insight in the estimation process. In this case, we also propose two different ways of choosing a proper procedure: those obtaining a mean square error (MSE) close to the true value, i.e. in the $[0.9\cdot\text{MSE},1.1\cdot\text{MSE}]$ interval (GI), or those that better correct the overestimation. In Figure \ref{scheme_options} we give some guidelines about the best options in terms of these four criteria considering different dependence and/or scale configurations. It is considered if there exists some dependence relation (Yes/No) and if there are covariates in different scales (Yes/No). Besides, we distinguish if these algorithms only perform properly in the $p>n$ or $n>p$ framework and what $\mathbf{X}^{\text{r}}$ or $\mathbf{X}^{\text{us}}$ version should be employed.

\section*{Funding Declaration}

This work was supported by the Conseller\'ia de Cultura, Educaci\'on e Ordenaci\'on Universitaria along with the Consellería de Econom\'ia, Emprego e Industria of the Xunta de Galicia under Project ED481A-2018/264; and MICIU/AEI/10.13039/501100011033 and the Competitive Reference Groups 2021–2024 (ED431C 2021/24) from the Xunta de Galicia through the ERDF under Project PID2020-116587GB-I00. We also acknowledge to the Centro de Supercomputación de Galicia (CESGA) for computational resources.

\section*{Disclosure statement}

The authors report there are no competing interests to declare.


\bibliographystyle{apalike}
\bibliography{Libreria_LASSO}

\end{document}